\documentclass[10pt,twocolumn,superscriptaddress,aps,pra]{revtex4-2}

\usepackage{dcolumn}

\usepackage{color}
\usepackage{amsmath}
\usepackage{amssymb,amsfonts,latexsym,cancel} 
\usepackage{graphicx}
\usepackage{hhline}
\usepackage{mwe}
\usepackage{amsbsy}
\usepackage{textcomp}
\usepackage{commath}
\usepackage{comment} 
\usepackage{mathrsfs}
\usepackage{bm}
\usepackage{array} 
\usepackage{subcaption}
\usepackage[margin=0.9in]{geometry}
\usepackage{ragged2e}

\DeclareMathOperator{\csch}{csch}

\usepackage{babel}
\newcommand{\bea}{\begin{eqnarray}}
\newcommand{\eea}{\end{eqnarray}}

\newcommand{\be}{\begin{equation}}
\newcommand{\ee}{\end{equation}}

\numberwithin{equation}{section}

\makeatletter

\begin{document}

\author{S. Navarro-Obregón}
\affiliation{Departamento de Física Teórica, Atómica y Óptica, Universidad de Valladolid,\\
47011 Valladolid, Spain}

\author{J. Queiruga}
\affiliation{Department of Applied Mathematics, University of Salamanca, Casas del Parque 2}
\affiliation{Institute of Fundamental Physics and Mathematics, University of Salamanca,\\
Plaza de la Merced 1, 37008 - Salamanca, Spain}

\title{Impact of the internal modes on the sphaleron decay}

\begin{abstract}
We study the sphaleron solutions in two deformations of the $\phi^6$ model and analyze the oscillons originated from them. We find that the presence of internal modes plays a crucial role in the sphaleron collapse. The positive internal modes triggered by a 
squeezing of the sphaleron are able to change the direction of collapse. We provide an analytical understanding behind this phenomenon.   
\end{abstract}
\maketitle

%============================================
%============================================
%============================================
%============================================
%============================================

\section{Introduction}

Sphalerons are localized, static, but unstable solutions of classical field equations that can be identified with saddle points in  configuration space \cite{Manton, Manton2}. These objects have been found in one or more spatial dimensions \cite{MSam, KS, Adam} in theories with or without topological solitons, and have revealed a crucial role in the baryon asymmetry in the context of electroweak theory \cite{KM, Klinkhamer, James}. 

 The sphaleron instability can be traced back to the presence of a negative internal mode in the linear perturbations. Excitations of its unstable mode may lead to its decay to a localized, oscillating, and long-lived solution called oscillon \cite{Dashen, Kudryavtsev, Bogolyubskii}. Oscillons owe their longevity to probe the nonlinearities of the theory \cite{Copeland}, and have found applications in many scenarios in theoretical physics, from dark matter \cite{Olle, Kawasaki, Arvanitaki} to cosmology \cite{Hindmarsh, Gorghetto, Blanco}. 

In spite of being two distinct objects, it has been suggested a possible relation between them \cite{MR}. This not only means that the profiles of oscillons are bounded by the corresponding sphaleron, but that the dynamics of the large amplitude oscillon arising in the decay of a sphaleron may be accurately captured by a collective coordinate model (CCM) based on sphaleron degrees of freedom (d.o.f). This was originally observed in the case of $\phi^3$ model \cite{MR}, the simplest theory allowing for both sphaleron and oscillon, where the pertinent d.o.f. are the amplitudes of the unstable and the massive mode of the sphaleron. This relation has been recently investigated in cases where the sphaleron does not support any positive energy bound mode \cite{Jose}. In that work, the authors introduced the first Derrick mode to explain some features of the oscillon structure emerging after the decay of the sphaleron, extending the possible oscillon-sphaleron relation for that regime. 

In this work, we aim to analyze the impact of the internal d.o.f of the sphaleron on its subsequent decay. In order to do that, we have proposed two families of models which are deformations of the $\phi^6$ theory in the same fashion as in Ref. \cite{Bazeia}. Both support analytical sphalerons, but differs as far as the spectral structure is concerned. In the first deformation, which we call the \textit{barrier model}, the sphalerons do not support any bound mode. In the second deformation, which we call the \textit{well model}, the sphaleron has a growing number of positive oscillating modes depending on the value of an adjustable parameter. We will show that the presence of positive internal modes in the sphaleron spectrum will play a crucial role in the dynamics of the decay.

The organization of this manuscript is as follows: in Section \ref{sec:models} we propose two deformations of the $\phi^6$ model and analyze their sphaleron solutions. In Section \ref{sec:mode_structure} we study the linear fluctuation spectrum of excitations on the background of the sphalerons. In Section \ref{sec:decay} we comment on the oscillon formation from sphalerons. In Section \ref{sec:CCM} we analyze the impact of the internal modes in the possible decay channels in both models. Finally, Section \ref{sec:summary} is devoted to our conclusions.   

%============================================
%============================================
%============================================
%============================================
%============================================

\section{The $\phi^6$ family of models}\label{sec:models}

\begin{figure}[!ht]
\centering
\begin{subfigure}[b]{\columnwidth}
  \centering
  \includegraphics[width=0.93\linewidth]{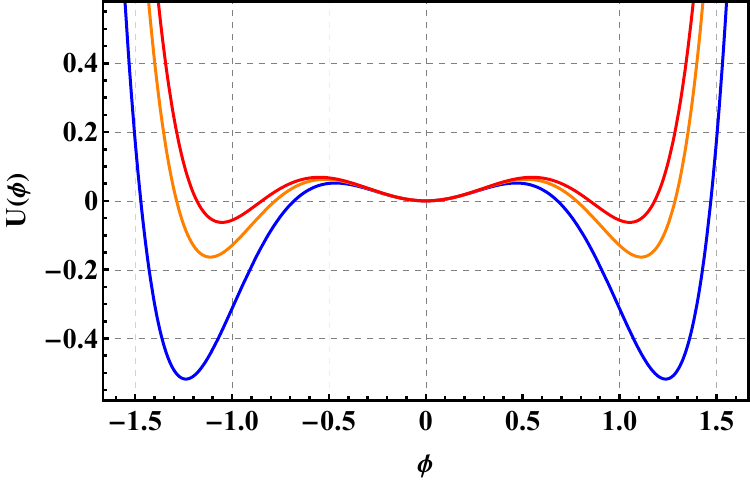}
  \caption{\small $s = 0.5$ (lower), $s = 0.7$ (middle), $s = 0.9$ (upper)\,.}
  \label{fig:potential_barrier}
\end{subfigure}

\vspace{0.4cm}

\begin{subfigure}[b]{\columnwidth}
  \centering
  \includegraphics[width=0.93\linewidth]{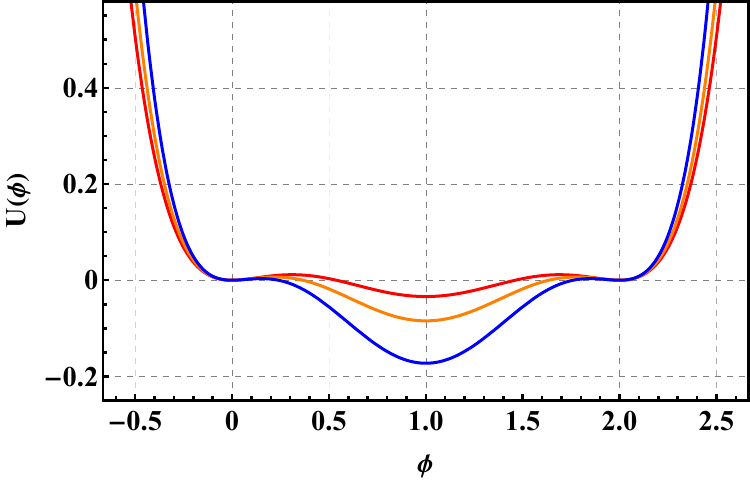}
  \caption{\small $s = 0.5$ (upper), $s = 0.75$ (middle), $s = 1.0$ (lower)\,.}
  \label{fig:potential_well}
\end{subfigure}
\caption{\small \justifying Potential of the barrier model (upper panel) and of the well model (lower panel) for different values of $s$.}
%\label{fig:potential}
\end{figure}

We introduce the following family of $\phi^6$ models % form and that are deformed by the presence of a parameter $s > 0$
\begin{equation}\label{eq:Action}
L = \int_{-\infty}^{\infty} \left( \dfrac{1}{2}\phi_t^2 - \dfrac{1}{2}\phi_x^2 - U_{6}(\phi;s)\right)\,dx,
\end{equation}
where the potential $U_{6}(\phi;s)$, which depends on a parameter $s$, is defined by  
\begin{equation}\label{potential_Barrier}
U_{6}(\phi;s) = \dfrac{1}{2}\phi^2(\tanh s - \phi^2)(\coth s - \phi^2),\,\, s>0.
\end{equation}
We will call this deformation the barrier model for reasons that will become clear later. For any finite value of $s$ the potential develops a false vacuum at $\phi=0$ and two symmetric true vacua. In the limit when $s$ goes to infinity one recovers the standard $\phi^6$ model with three vacua located at $\phi=0,\, \pm 1$. We depict the potential (\ref{potential_Barrier}) for different values of $s$ in Fig. \ref{fig:potential_barrier}. The corresponding equation of motion reads as
\begin{equation}\label{eq:barrier}
\square \phi + \phi - 4\coth 2s \, \phi^3 + 3\,\phi^5 = 0\,.
\end{equation} 
The presence of the false vacuum allows for unstable finite-energy solutions of (\ref{eq:barrier}) called sphalerons which interpolate between the false vacuum and the zero of the potential. They are also piece-wise solutions of the following first order (BPS) equations \cite{Bazeia, Alberto, Manton3}   
\begin{equation}\label{eq:BPS}
\phi'(x)=\pm \sqrt{2\, U_{6}(\phi;s)}\,.
\end{equation}
This equation can be easily integrated. Its solutions are the sphalerons of the model
\begin{equation}\label{Sphaleron_barrier}
\phi_S(x;s) = \pm \sqrt{\dfrac{\sinh 2s }{\cosh 2s + \cosh 2x}}\,.
\end{equation}

\begin{figure}[!ht]
\centering
\begin{subfigure}[b]{\columnwidth}
  \centering
  \includegraphics[width=0.92\linewidth]{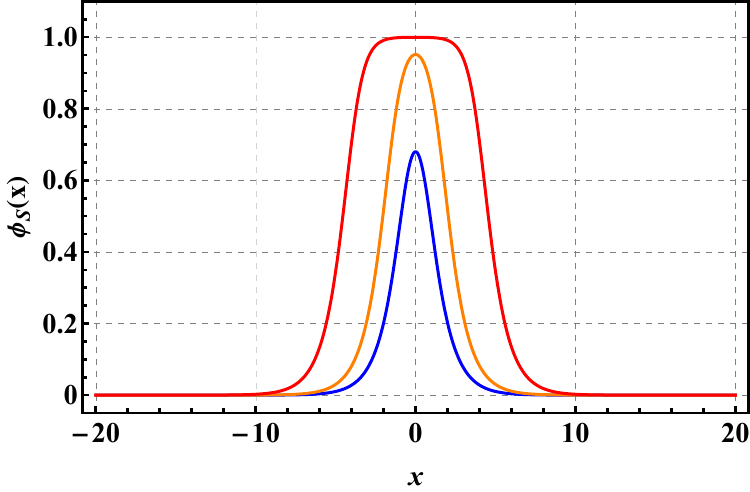}
  \caption{\small $s = 0.5$ (lower), $s = 1.5$ (middle), $s = 4.0$ (upper)\,.}
  \label{fig:sphaleron_barrier}
\end{subfigure}

\vspace{0.3cm}

\begin{subfigure}[b]{\columnwidth}
  \centering
  \includegraphics[width=0.91\linewidth]{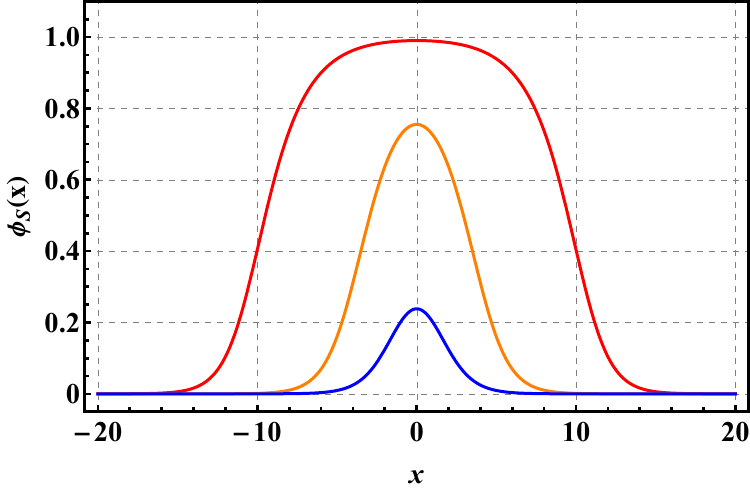}
  \caption{\small $s = 0.01$ (upper), $s = 0.25$ (middle), $s = 1.0$ (lower)\,.}
  \label{fig:sphaleron_well}
\end{subfigure}
\caption{\small \justifying Sphaleron profile in the barrier model (up) and in the well model (down) for different values of $s$.}
\label{fig:sphlaleron}
\end{figure}

For small $s$ the profile is lump-like, and it gets wider as $s$ increases (see Fig. \ref{fig:sphlaleron} (a)), resembling a kink-antikink (KAK) pair. The energy of the sphaleron can be computed analytically 
\begin{equation}\label{eq:ener_bar}
    E(s)=\frac{1}{4}\csch^2(2s)\left(\sinh(4s)-4s\right).
\end{equation}
We learn from (\ref{eq:ener_bar}) that the energy of the sphaleron grows with $s$ and it is bounded by the energy of the KAK pair of the underlying $\phi^6$ model, that is $E(s)<1/2$.

Another possible deformation consists in shifting the central vacuum so that it becomes the new true vacuum. This new potential reads as
\begin{equation}\label{potential_Well}
U_{s6}(\phi) = \dfrac{(\phi - 1)^2 - \tanh^2 s}{8(1 -\tanh^2 s)}\left((\phi - 1)^2 - 1\right)^2,
\end{equation}
and we have called it well model. The further displacement to the right will be convenient later on when comparing both deformations. This new potential has two false vacua at $\phi \in \{0,\,2\}$, while at $\phi=1$ has a true vacuum. Now, the standard $\phi^6$ potential rescaled by a factor $4$ and shifted to the right is recovered in the limit $s\rightarrow 0$ (the reason for this rescaling is to fix the spectrum mass threshold at $\omega^2=1$). The profile of this potential for different values of $s$ is presented in Fig. \ref{fig:potential_well}. The associated equation of motion is
\begin{eqnarray}\label{eq:well}
&& \hspace{-0.3cm} \square \phi + \phi - \dfrac{3}{2}\left(2 + \cosh 2s \right)\phi^2 + (\dfrac{7}{2} + 3 \cosh 2s)\phi^3\nonumber\\
&-& \dfrac{15}{4 (1 - \tanh^2 s)}\phi^4 + \dfrac{3}{4 (1 - \tanh^2 s)}\phi^5 = 0.
\end{eqnarray}
Again, the sphalerons are piece-wise solutions of (\ref{eq:BPS}). They have the following form 
\begin{equation}\label{Sphaleron_well}
\phi_{S}(x;s) = 1 \pm 2\,g(x;s)\sinh s\,,
\end{equation}
where
\begin{equation}\label{eq:g_function}
g(x;s) = \dfrac{\cosh x/2}{\sqrt{3 + \cosh 2s + 2\cosh x \sinh^2 s}}\,.
\end{equation}

Now, the sphaleron profile resembles a KAK pair as $s$ decreases (see Fig. \ref{fig:sphlaleron} (b)). The energy can be also computed analytically although the explicit expression is not particularly illuminating. The energy decreases with $s$ and is bounded from above by the energy of the corresponding KAK pair, that is $E(s)<1/4$ (note that in the limit $s\rightarrow 0$ the standard $\phi^6$ model is obtained but rescaled by a factor $4$). 

A first difference with previously studied models is that, unlike the $\phi^3$ model \cite{MR} or the inverted $\phi^4$ model \cite{Jose}, these families of potentials are bounded from below, avoiding the field becoming singular. As we will show in the following section, the most notorious difference between the aforementioned models will lie in the number of shape modes supported by the corresponding sphalerons.

%============================================
%============================================
%============================================
%============================================
%============================================

\section{The mode structure}\label{sec:mode_structure}

The sphaleron, like any topological soliton, can host internal modes \cite{MR, Alberto}. 
To compute the linear spectrum around the sphaleron solution, we add as usual a small perturbation in the following way
\begin{equation}\label{perturbation}
\phi(x,t;s) = \phi_{S}(x;s) + \eta(x;s)e^{i \omega t}.
\end{equation}

Inserting (\ref{perturbation}) into the field equation and expanding up to linear order we get 
\begin{equation}
\left[ - \dfrac{d^2}{dx^2} + U_6''\left(\phi_{S}(x;s)\right)\right]\eta(x;s) = \omega^2(s)\,\eta(x;s).
\end{equation}

A relevant feature of the sphaleron in the barrier model (\ref{potential_Barrier}) is that it only possesses a negative unstable mode and a zero mode in the discrete part of the linear spectrum, but there are no positive bound modes. This can be explained qualitatively by noticing that, at least for large values of $s$, the sphaleron is like a KAK pair $(0,1)+(1,0)$ in the $\phi^6$ model. Such a pair is known not to support bound modes \cite{Dorey}. The reason is that the inner vacuum at $\phi = 1$ has a higher mass threshold than the outer vacuum at $\phi = 0$. Therefore, in the spectral problem, the effective potential has a barrier in the middle of the constituent kinks (this is the reason behind the denomination \textit{barrier model}).

\begin{figure}[!ht]
\centering
\hspace{-0.6cm}
\begin{subfigure}[b]{\columnwidth}
  \centering
  \includegraphics[width=0.92\linewidth]{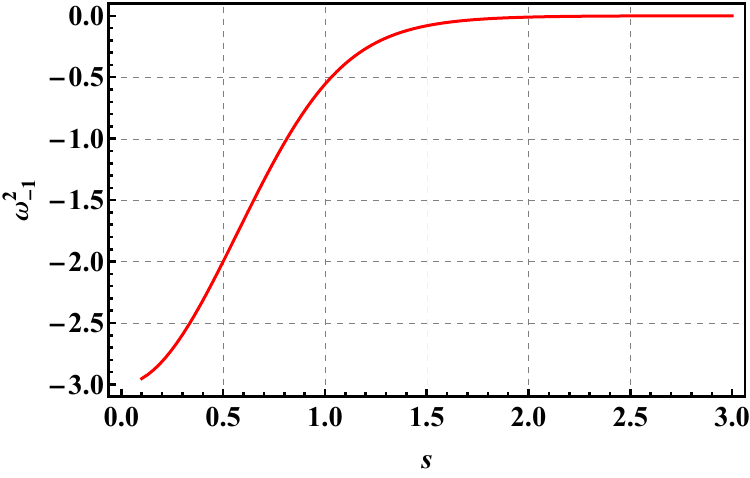}
  \caption{\small Barrier model unstable mode\,.}
  \label{fig:unstable_barrier}
\end{subfigure}

\vspace{0.4cm}

\begin{subfigure}[b]{\columnwidth}
  \centering
  \hspace{-0.6cm}
  \includegraphics[width=0.92\linewidth]{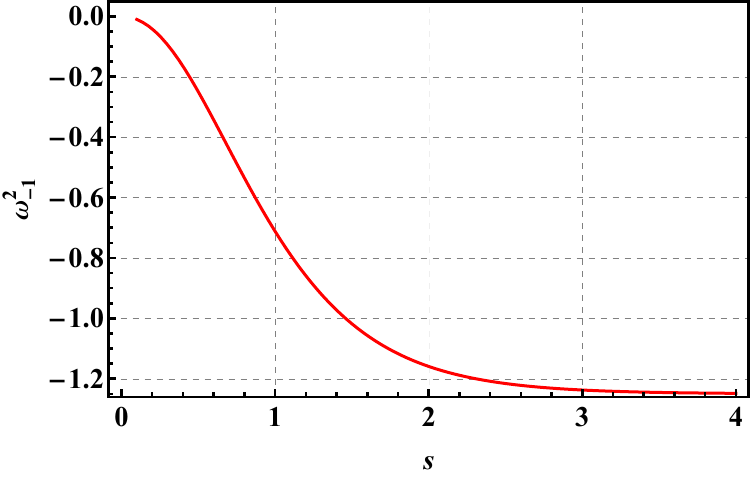}
  \caption{\small Well model unstable mode\,.}
  \label{fig:unstable_well}
\end{subfigure}
\caption{\small Dependence of the unstable mode on $s$.}
\label{fig:Unstable_mode}
\end{figure}

We present the dependence of the unstable mode $\omega^{2}_{-1}$ on the parameter $s$ in Fig. \ref{fig:Unstable_mode} a. As $s$ grows the sphaleron resembles a KAK pair of the standard $\phi^6$ model. As a consequence, the unstable mode approaches the zero mode. In the limit $s \rightarrow \infty$ the modes degenerate, corresponding with two zero modes on top of the sphaleron walls, that is, the constituent subkinks. 

\begin{figure}[!ht]
\centering
{
\includegraphics[width=0.93\columnwidth]{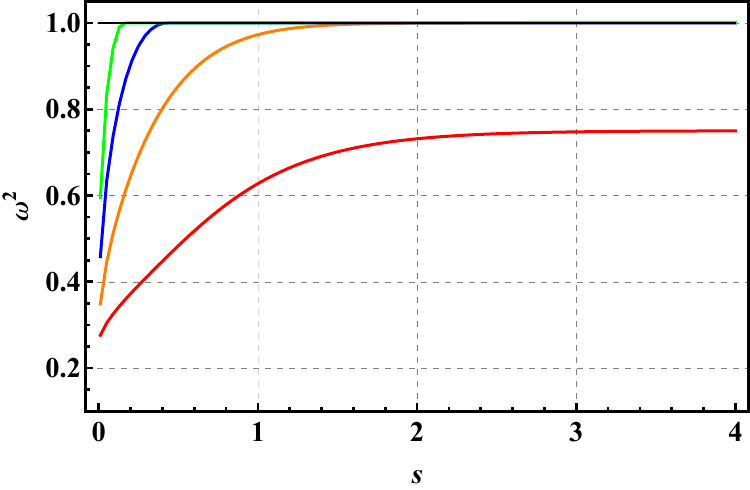}   
}
\caption{\small \justifying First four massive bound modes of the sphaleron in the well model (\ref{potential_Well}).}
\label{fig:modeStructure}
\end{figure}

On the other hand, unlike the sphaleron in the barrier model, the sphaleron in the well model (\ref{potential_Well}) can host shape modes, increasing in number as $s \rightarrow 0$. This can be seen again from the KAK pair picture. Now, as $s \rightarrow 0$, we find a more and more separated pair of antikink-kink $(0,1)+(1,0)$ in the $\phi^6$ model. Note that due to the shifting of the vacua, this configuration actually corresponds to the $(1,0)+(0,1)$ pair of the original $\phi^6$ model \cite{Dorey}. Such a pair supports an effective potential in the linear problem which leads to a growing number of modes as the distance between the pair grows, since it exhibits a well in the middle of the constituent kinks (this is the reason behind the denomination \textit{well model}). The structure of the first lower positive bound modes is depicted in Fig. \ref{fig:modeStructure}. It can be verified that the positive bound modes held by the sphaleron for small values of $s$ resemble the spatial profiles of the delocalized modes appearing in the $\phi^6$ KAK sector as suggested before \cite{Dorey, Garcia}.

An analytical study of the asymptotic eigenvalue problem for $\omega^2 = 1$ shows that the eigenstate at the threshold has an increasing number of nodes for decreasing values of $s$
\begin{equation}
\eta_{\omega^2 = 1} \propto \text{I}\left[ 0,\dfrac{2\, e^{-x/2}\sqrt{-12 - 48e^{2s} - 12e^{4s}}}{\sqrt{1 - 2e^{2s} + e^{4s}}}\right],
\end{equation}
where $I[n,x]$ denotes the modified Bessel function of the first kind. This function has an unbounded number of nodes as $s$ decreases. From the node theorem in Sturm-Lioville theory it follows that the number of modes below the threshold $\omega^2=1$ is also unbounded.

As we will see in the following sections, the presence of internal modes introduces new mechanisms which play a crucial role in the sphaleron decay. 

%============================================
%============================================
%============================================
%============================================
%============================================

\section{Sphaleron decay}\label{sec:decay}

The presence of a negative mode in the sphaleron spectrum is at the origin of the sphaleron instability. Thus, the sphaleron can decay when perturbed slightly. Particularly, it decays in two different channels: one leads to the creation of an accelerating KAK pair (associated with the true vacuum), while the other results in the formation of an oscillon (linked to the false vacuum). The initial condition for an slightly perturbed sphaleron with its unstable mode is
\begin{eqnarray}\label{IC_decay}
\phi(x,0) &=& \phi_{S}(x;s) \pm A\,\eta_{-1}(x;s)\,,\nonumber \\
\dot{\phi}(x,0) &=& 0\,.
\end{eqnarray}
Here $\eta_{-1}(x;s)$ represents the sphaleron unstable mode, and $A$ is its initial amplitude. Moreover, the plus sign denotes a perturbation in the expanding direction, whereas the minus sign accounts for an excitation in the collapsing direction. Note that for a big sphaleron the negative mode can be understood as a symmetric combination of the zero modes of the constituent subkinks. 

As aforementioned, a perturbation in the expanding direction gives rise to an accelerating KAK pair. This acceleration can be understood through an energy conservation argument: as the maximum of the sphaleron descends towards the true minima, the field reaches negative values of the potential. As the sphaleron grows,  a plateau of negative potential energy forms at its center. This negative energy has to be compensated by an increase of the kinetic energy at the sphaleron walls.

\begin{figure}[!ht]
\centering
\begin{subfigure}[b]{\columnwidth}
  \centering
  \includegraphics[width=0.91\linewidth]{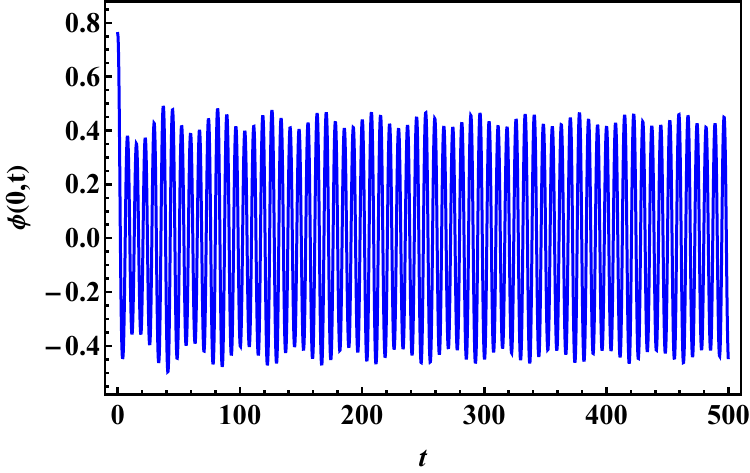}
  \caption{\small Barrier model: $s = 1.267$, $A = 0.6$\,.}
  \label{fig:decay_barrier_1}
\end{subfigure}

\vspace{0.3cm}

\begin{subfigure}[b]{\columnwidth}
  \centering
  \includegraphics[width=0.91\linewidth]{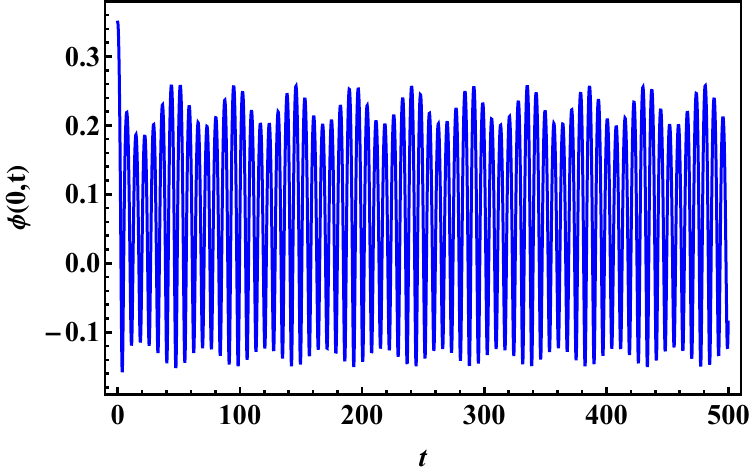}
  \caption{\small Well model: $s = 0.45$, $A = 0.6$\,.}
  \label{fig:decay_well_1}
\end{subfigure}
\caption{\small \justifying Decay of the sphaleron in the barrier model (upper panel) and in the well model (lower panel). The figures show the field at the origin $\phi(0,t)$. The frequency of the unstable mode is $\omega^2_{-1} \approx -0.2056$.}
\label{fig:decay_comparison_low}
\end{figure}

\begin{figure*}
    \begin{subfigure}[b]{0.5\textwidth}
        \centering
        \includegraphics[width=0.94\columnwidth]{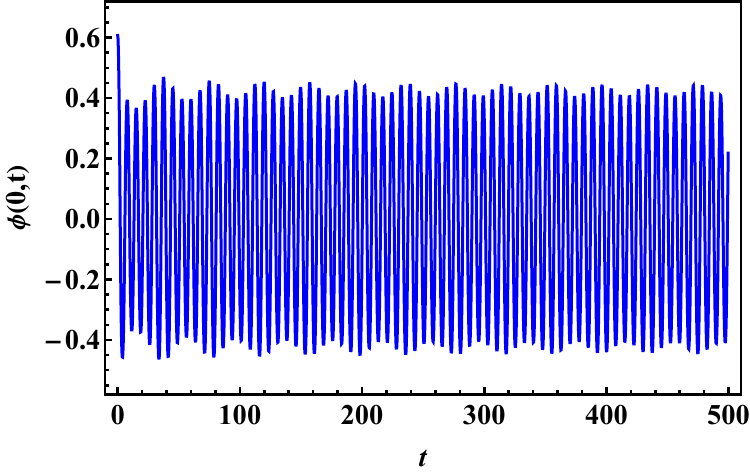}
        \caption{\small Barrier model: $s = 0.75$, $A = 0.3$\,.}
        \label{fig:decay_barrier_2}
    \end{subfigure}%
    \begin{subfigure}[b]{0.5\textwidth}
        \centering
        \includegraphics[width=0.96\columnwidth]{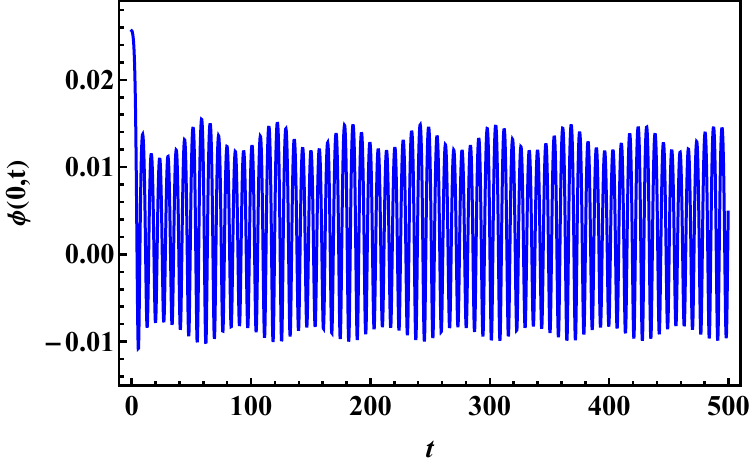}
        \caption{\small Well model: $s = 2.16$, $A = 0.001$\,.}
        \label{fig:decay_well_2}
    \end{subfigure}
    \caption{\small \justifying Decay of the sphaleron in the barrier model (left panel) and in the well model (right panel). The figures show the field at the origin $\phi(0,t)$. The frequency of the unstable mode is $\omega^2_{-1} = - 1.183$.}
    \label{fig:decay_comparison_high}
\end{figure*}

Regarding the collapsing direction, there have been selected appropriate values of $s$ so that the unstable frequencies $\omega^2_{-1}$ are equal in both deformations of the $\phi^6$ model, enabling us to compare the decays of their respective sphalerons. When examining the decay of the sphalerons,
we initially observe a distinguishable double quasi-periodic structure in both models for suitable initial
amplitudes, which depend on the size of the sphaleron. Plots of the decay processes in the barrier and well model for different $\omega^2_{-1}$ can be found in Fig. \ref{fig:decay_comparison_low} (for a low negative frequency case) and in Fig. \ref{fig:decay_comparison_high} (for a high negative frequency case). For lower or higher values of the amplitude, we initially observe either a long-living oscillon with imperceptible modulation that becomes noticeable at later times, or a fast-decaying oscillon, respectively. This may suggest that the double quasi-periodic structure is an intermediate state between stable and unstable oscillons \cite{Ghersi}, although a complete understanding of the nature of this internal degree of freedom is still lacking \cite{Blaschke:2024uec}.

As shown in \cite{Jose}, the oscillon formed from the decay of the sphaleron does present a modulated amplitude even though the original sphaleron does not hold any shape mode. The analogy of the sphaleron as constituted by a KAK pair may indicate that the evolution of the oscillon is influenced by the dynamics of the individual kinks. However, we have seen here that the double quasi-periodic structure emerges even if the constituting kinks do not hold positive bound modes. This suggests that the influence of the constituent kinks in the subsequent oscillon could be manifest in an extremely nontrivial way, and needs further investigation.

\begin{figure}[!ht]
\centering
\begin{subfigure}[b]{\columnwidth}
  \centering
  \includegraphics[width=0.86\linewidth]{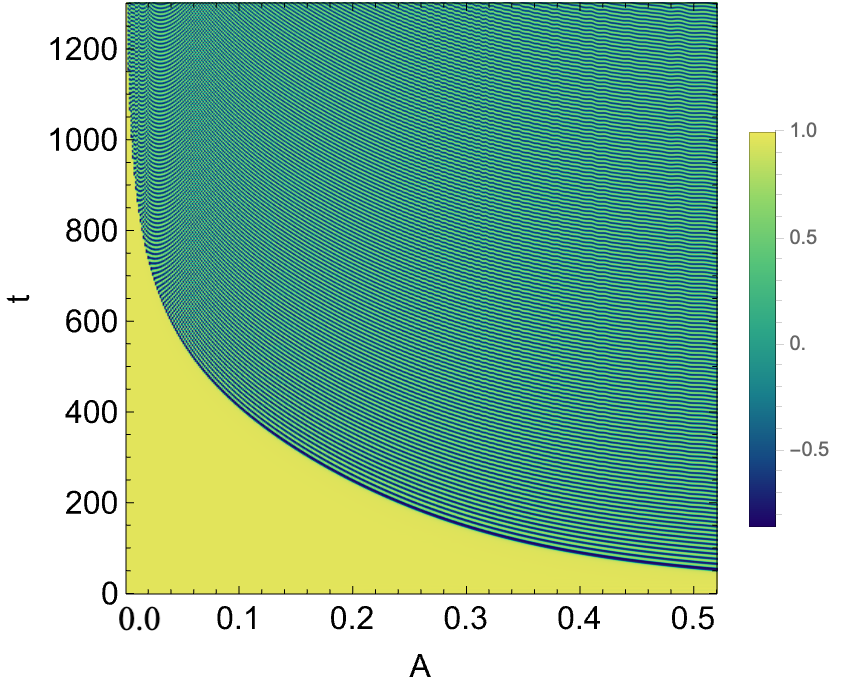}
  \caption{\small Barrier model: $s = 3.5$\,.}
  \label{fig:decay_barrier_large}
\end{subfigure}

\vspace{0.3cm}

\begin{subfigure}[b]{\columnwidth}
  \centering
  \includegraphics[width=0.86\linewidth]{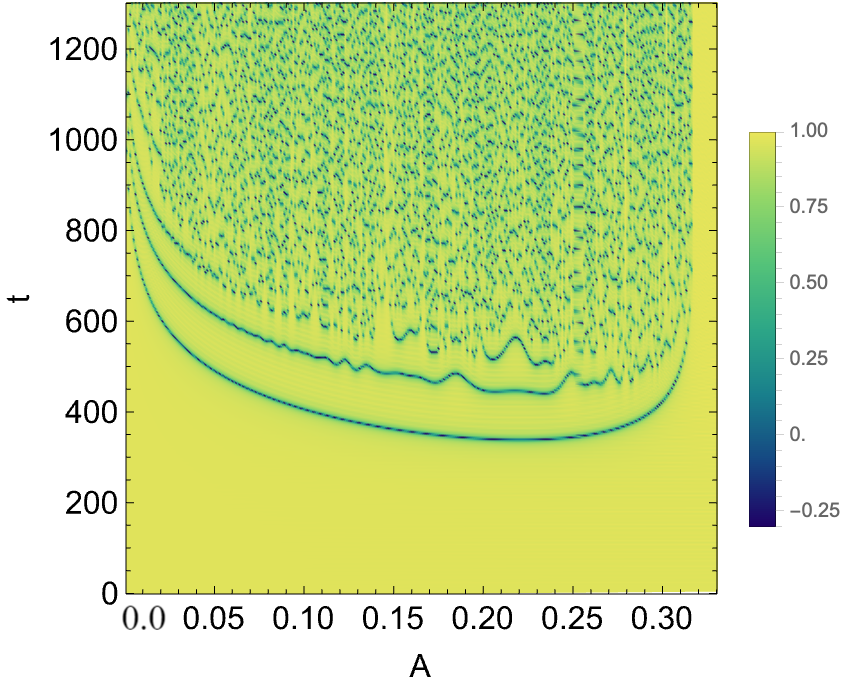}
  \caption{\small Well model: $s = 0.008$\,.}
  \label{fig:decay_well_large}
\end{subfigure}
\caption{\small \justifying Comparison of the decay of the sphaleron in both the barrier model (upper panel) and the well model (lower panel) for an established KAK pair. The color palette show the field at the origin $\phi(0,t)$.}
\label{fig:CriticalPhenomenon}
\end{figure}

In the barrier model, the subsequent oscillon after the  decay is quite insensitive to the initial size of the sphaleron (measure by the $s$ parameter). However, in the well model, the sphaleron size determines the number of internal modes and this has a direct impact on the decay. As representative values of the parameter $s$, we consider $s = 3.5$ for the barrier model and $s = 0.008$ for the well model. Within this range of values, the sphaleron can be seen as a well-defined KAK pair. The evolution of the sphaleron for various initial amplitudes of the unstable mode is illustrated in Fig. \ref{fig:CriticalPhenomenon}. The color palette in the figure represents the value of the field at the origin.
\\
\indent First, one can observe that the sphaleron decays earlier in both models as the amplitude A of the unstable mode increases. However, it can be distinguished a more chaotic pattern in the well model. Here, after the collapse, the sphaleron bounces and reappears again. This behavior can be attributed to the presence of internal modes, which are able to store energy for certain time and transfer it back to the constituent subkinks leading to a growth of the solution. Only the even parity modes play the leading role in this process. This will be seen easily from the perspective of an effective Lagrangian that we will build later on in Section \ref{sec:CCM}. 

Remarkably, a novel phenomenon emerges for sufficiently large amplitudes and small values of $s$ in the well model. Above some critical value of the amplitude of the unstable mode, the sphaleron expands as two outgoing accelerating kinks instead of collapsing. In Fig. \ref{fig:decay_well_large} the critical amplitude is $A_{crit} \approx 0.3168$. We will be explored this in more detail in Section \ref{sec:CCM}.

%============================================
%============================================
%============================================
%============================================
%============================================

\section{CCM based on sphaleron d.o.f}\label{sec:CCM}

In this section we aim to describe the sphaleron dynamics when its unstable mode has been triggered. In order to do that, we shall build an effective model based on sphaleron d.o.f. through CCM approach. In this scheme, the effective theory is computed by introducing a field configuration ansatz $\Phi(x; X^{i}(t))$ into the Lagrangian density and integrating over the spatial coordinate, obtaining a mechanical model of the form \cite{Oles,Andrzej}

\begin{equation}
L[X^i] = \dfrac{1}{2}g_{ij}\dot{X}^{i}\dot{X}^{j} - V_{eff}(X^{i}),
\end{equation}
with the metric on the moduli space and the effective potential given by
\begin{eqnarray}
g_{ij} &=& \int_{\mathbb{R}} \dfrac{\partial \Phi}{\partial X^i}\dfrac{\partial\Phi}{\partial X^j}\,dx,\\ V_{eff}(X^{i}) &=& \int_{\mathbb{R}} \left(\dfrac{1}{2}\left(\dfrac{\partial \Phi}{\partial x}\right)^2 + V(\Phi) \right)\, dx.
\end{eqnarray}

\noindent
Here $X^i(t)$ are time-dependent variables representing global excitations that encode the dynamics of the soliton.

First, we will study the evolution of the oscillon formed after the collapse of the sphaleron. Then, we will explore the formation of the KAK pair in the well model for large enough values of the unstable mode amplitude, and the mechanism behind this phenomenon will be explained.

%============================================
%============================================
%============================================

\subsection{CCM in the barrier model}

When dealing with the sphaleron in the barrier model (\ref{potential_Barrier}), we have shown that it does not hold internal positive modes (see Section \ref{sec:mode_structure}). As shown in \cite{Jose}, it is still possible to build an effective model based on the unstable mode and the Derrick mode. The Derrick mode describes a scaling deformation and it is defined by the following
expansion
\begin{equation}
    \phi_S(x(1+\epsilon);s)=\phi_S (x,s)+\epsilon\, \eta_D(x;s)+...
\end{equation}
A naive ansatz based on the sphaleron d.o.f would look as follows
\begin{eqnarray}\label{effective_barrier}
\Phi(x,t) &=& \phi_{S}(x;s)\nonumber\\
&+& a(t)\eta_{-1}(x;s) + b(t)\eta_{D}(x;s).
\end{eqnarray}
Nevertheless, it is expected that, for high values of $s$, the field configuration $(\ref{effective_barrier})$ does not capture the evolution of the oscillon since, due to the emergence of the underlying KAK structure, the unstable mode $\eta_{-1}(x;s)$ will exhibit two separated peaks around the positions of the constituent subkinks, that is, there will be no overlap between the mode and the oscillon. As a consequence, the unstable mode will not be a good degree of freedom in the description of the oscillon. However, as we will show, it is still possible to describe properly the oscillon dynamics with an adapted model based on sphaleron d.o.f.

\begin{table}[!ht]
\centering
\renewcommand{\arraystretch}{1.5}
\setlength{\tabcolsep}{12pt}
\begin{tabular}{|c|c|c|c|}
\hline
\hline
\textbf{$s$} & \textbf{$\omega_{osc}^{ET(O)}$} & \textbf{$\omega_{osc}^{ET(S)}$} & \textbf{$\omega_{osc}^{FT}$} \\
\hline
\hline
0.1 & 0.881 & 0.88 & 0.881 \\
\hline
1.267 & 0.805 & 0.817 & 0.813 \\
\hline
\end{tabular}
\caption{\small \justifying Oscillation frequencies for $s = 0.1$ and $s = 1.267$ in field theory and in the effective models for the barrier model. The upper index O or S denotes the oscillon-based effective model or the sphaleron-based effective model.}
\label{tab:oscillation}
\end{table}

\begin{table}[!ht]
\centering
\renewcommand{\arraystretch}{1.5}
\setlength{\tabcolsep}{12pt}
\begin{tabular}{|c|c|c|c|}
\hline
\hline
\textbf{$s$} & \textbf{$\omega_{mod}^{ET(O)}$} & \textbf{$\omega_{mod}^{ET(S)}$} & \textbf{$\omega_{mod}^{FT}$} \\
\hline
\hline
0.1 & 0.124 & 0.125 & 0.114 \\
\hline
1.267 & 0.225 & 0.238 & 0.185 \\
\hline
\end{tabular}
\caption{\small \justifying Modulation frequencies for $s = 0.1$ and $s = 1.267$ in field theory and in the effective models for the barier model. The upper index O or S denotes the oscillon-based effective model or the sphaleron-based effective model.}
\label{tab:modulation}
\end{table}

Let us first introduce an effective model derived from the oscillon itself \cite{Navarro,Fodor:2008es}. As shown in \cite{Jose}, an \textit{ad hoc} effective model to describe the subsequent oscillon formed from the sphaleron decay may contain the leading profile of the oscillon from a small amplitude expansion \cite{Fodor} and a second contribution accounting for the associated Derrick mode
\begin{equation}\label{eq:oscillon_proposal}
\Phi(x;a,b) = \dfrac{a(t)}{\cosh(x \lambda)} + b(t)\dfrac{x\tanh(x \lambda)}{\cosh(x\lambda)}.
\end{equation}
Here, $a(t)$ and $b(t)$ are collective coordinates, and $\lambda$ is a parameter measuring the size of the oscillon. 

\begin{figure*}[!ht]
    \begin{subfigure}[b]{0.331\textwidth}
        \centering
        \includegraphics[width=0.99\columnwidth]{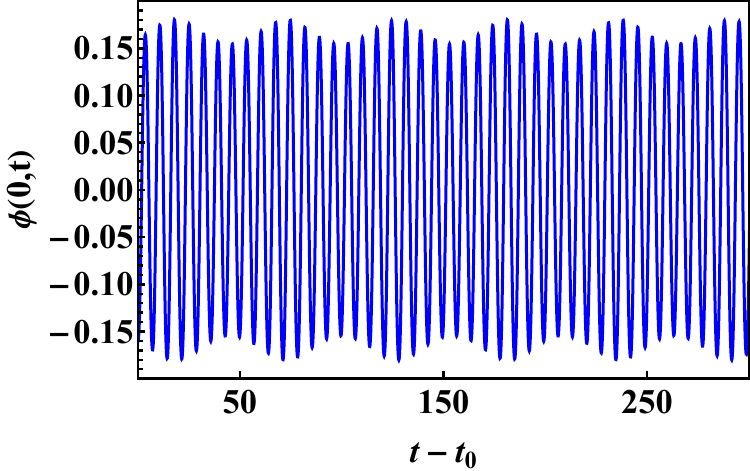}
        \caption{\small Field theory\,.}
        \label{Decay_Barrier_s01_Comparison}
    \end{subfigure}%
    \begin{subfigure}[b]{0.331\textwidth}
        \centering
        \includegraphics[width=0.99\columnwidth]{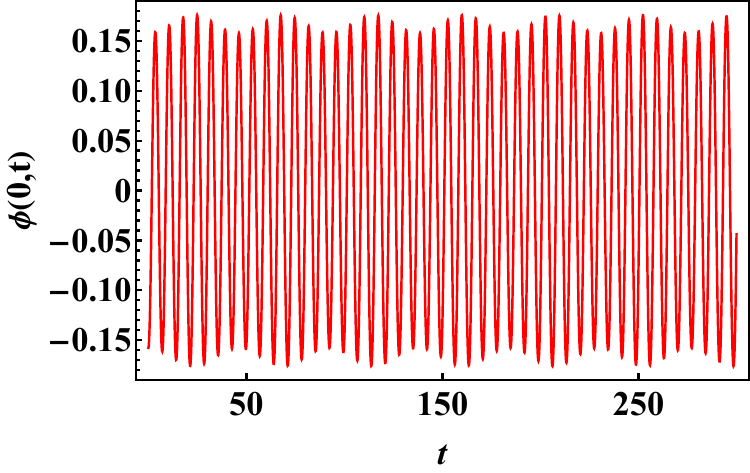}
        \caption{\small Oscillon-based effective model\,.}
        \label{LeastSquare_Barrier_01}
    \end{subfigure}
    \begin{subfigure}[b]{0.331\textwidth}
        \centering
        \includegraphics[width=0.99\columnwidth]{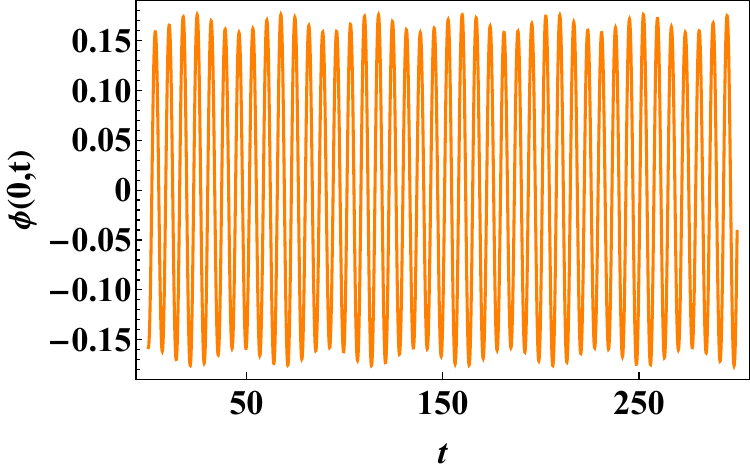}
        \caption{\small Sphaleron-based effective model\,.}
        \label{LeastSquare_Barrier_01_New}
    \end{subfigure}
    \caption{\small \justifying Comparison between field theory dynamics (left panel), and the least squares fitting in the oscillon-based (middle panel) and in the sphaleron-based model (right panel) for $s = 0.1$ in the barrier model.}
    \label{fig:comparison_barrier_01}
\end{figure*}

\begin{figure*}[!ht]
    \begin{subfigure}[b]{0.331\textwidth}
        \centering
        \includegraphics[width=0.99\columnwidth]{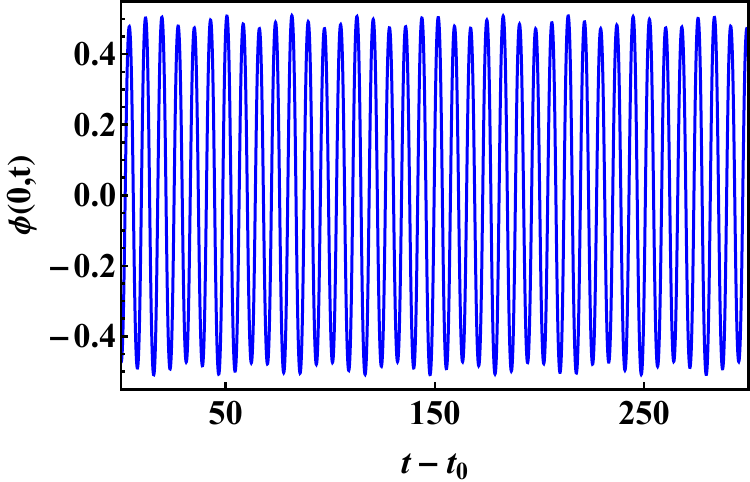}
        \caption{\small Field theory\,.}
        \label{Decay_Barrier_s1267_Comparison}
    \end{subfigure}%
    \begin{subfigure}[b]{0.331\textwidth}
        \centering
        \includegraphics[width=0.99\columnwidth]{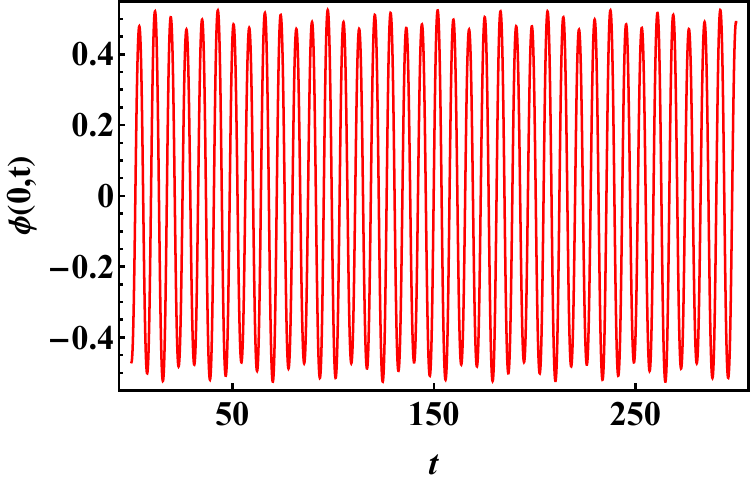}
        \caption{\small Oscillon-based effective model\,.}
        \label{LeastSquare_Barrier_126}
    \end{subfigure}
    \begin{subfigure}[b]{0.331\textwidth}
        \centering
        \includegraphics[width=0.99\columnwidth]{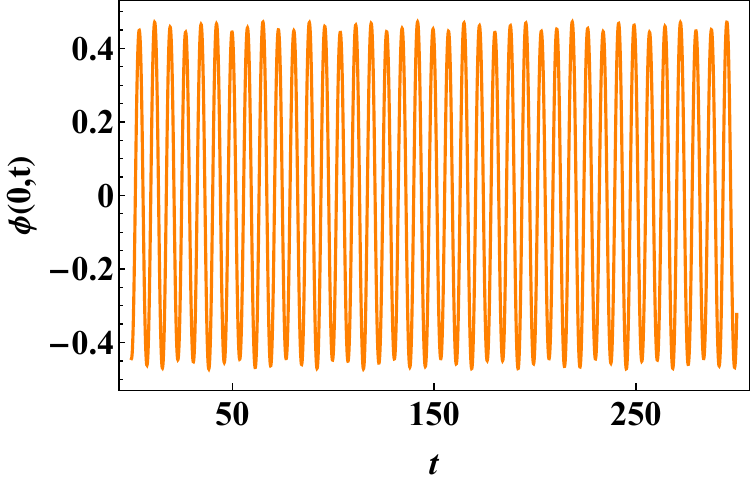}
        \caption{\small Sphaleron-based effective model\,.}
        \label{LeastSquare_Barrier_1267_New}
    \end{subfigure}
    \caption{\small \justifying Comparison between field theory dynamics (left panel), and the least squares fitting in the oscillon-based (middle panel) and in the sphaleron-based model (right panel) for $s = 1.267$ in the barrier model.}
    \label{fig:comparison_barrier_1267}
\end{figure*}

In our case, introducing (\ref{eq:oscillon_proposal}) into the Lagrangian density (\ref{eq:Action}) with the potential (\ref{potential_Barrier}) and integrating over the space we obtain the following metric on the moduli space
\begin{equation}\label{eq:metric}
g_{ij} = \begin{pmatrix}
    \dfrac{2}{\lambda} & \dfrac{1}{\lambda^2} \\
    \dfrac{1}{\lambda^2} & \dfrac{12 + \pi^2}{18\lambda^3} \\
\end{pmatrix}\,,
\end{equation}
and the following effective potential
\begin{eqnarray}
&&V_{eff}(a,b) = \frac{3 + \lambda^2}{3\lambda} a^2 + \frac{3 - \lambda^2}{3\lambda^2}ab + \frac{60+ \pi^2(5 + 7\lambda^2)}{180\lambda^3} b^2\nonumber\\
&& - \left(\frac{4}{3\lambda} a^4 + \frac{4}{3\lambda^2}a^3b + \frac{6\pi^2}{45\lambda^3}a^2b^2 \right.\nonumber \\
&& \left. - \frac{30 - 7\pi^2}{45\lambda^4}ab^3 - \frac{240 - 20\pi^2 - 3\pi^4}{900\lambda^5} b^4\right)\coth 2s\nonumber\\
&& + \frac{8}{15\lambda} a^6 + \frac{24}{45\lambda^2} a^5b - \frac{7 - 2\pi^2}{21\lambda^3}a^4b^2 \nonumber \\
&& - \frac{1 - \frac{10}{63}\pi^2}{\lambda^4} a^3b^3 - \frac{1260 + 5\pi^2 - 21\pi^4}{1890\lambda^5}a^2b^4 \nonumber\\ 
&& - \frac{630 + 475\pi^2 - 63\pi^4}{4725\lambda^6} ab^5\nonumber\\
&& + \frac{1}{\lambda^7}\left(\frac{1}{105} - \frac{13\pi^2}{378} + \frac{37\pi^4}{32400} + \frac{31\pi^6}{116424}\right)b^6.
\end{eqnarray}

Finally, the CCM is evolved using as initial conditions for $a(0),\, b(0)$ and $\lambda$ the optimal values obtained by fitting (\ref{eq:oscillon_proposal}) to the field configuration in field theory in an instant of time $t_0$ where the oscillon has settled down \cite{Jose}. Particularly, it is selected an instant of time which corresponds to a turning point, so that the kinetic energy can be neglected and, therefore, it is reasonable to assume $\dot{a}(0) = \dot{b}(0) = 0$. 

Two cases have been considered: $s = 0.1$ and $s = 1.267$. These examples represent two limits with significantly different unstable frequencies $\omega_{-1}^2$ and, consequently, different sizes. The initial configuration for field theory has been given by (\ref{IC_decay}) with the amplitudes $A = 0.001$ and $A = 0.1$ for the respective cases.

To evolve the CCM, the fitting has been performed at the turning points $t_0 = 7007.69$ and $t_0 = 7109.87$ respectively. The comparison between field theory and the effective theory based on the oscillon (\ref{eq:oscillon_proposal}) is shown in Fig. \ref{fig:comparison_barrier_01} and Fig. \ref{fig:comparison_barrier_1267} by representing the values of the field configuration at the origin $\phi(0,t)$. It is clearly visible that the effective model captures the amplitude modulation in both cases. The corresponding oscillation and modulation frequencies are listed in Table \ref{tab:oscillation} and Table \ref{tab:modulation}. These tables show a spectacular agreement in the oscillation frequency, with only a slight deviation appearing in the modulation frequency, being more noticeable for the $s = 1.267$ case.

This agreement suggests that one can describe the oscillon evolution with a degree of freedom accounting for the oscillon profile amplitude and an additional degree of freedom describing small changes in the oscillon size. To support this hypothesis, we suggest an oscillon profile entirely based on the sphaleron itself. 
We approximate now the oscillon profile by
\begin{eqnarray}
\Phi = a \sqrt{\dfrac{\sinh 2s }{\cosh 2s + \cosh 2 \lambda x}},
\end{eqnarray}
that is, we consider the sphaleron profile along with the fitting parameters $a$ and $\lambda$, accounting for the amplitude and the size respectively. Assuming a small scale deformation $\lambda \rightarrow \lambda + \epsilon$ and expanding up to first order, we obtain
\begin{eqnarray}
\Phi &=& a\sqrt{\dfrac{\sinh 2s }{\cosh 2s + \cosh 2 \lambda x}}\nonumber\\
&-& \epsilon a\dfrac{x\, \sqrt{\sinh 2s} \sinh 2\lambda x}{(\cosh 2s + \cosh 2\lambda x )^{3/2}}.
\end{eqnarray}
Finally, we promote the amplitudes to independent collective coordinates, so the previous expression reads  
\begin{eqnarray}\label{eq:sphaleron_proposal}
&&\Phi(x,t) = a(t)\sqrt{\dfrac{\sinh 2s }{\cosh 2s + \cosh 2 \lambda x}}\nonumber\\
&& + b(t)\dfrac{x\, \sqrt{\sinh 2s} \sinh 2\lambda x}{(\cosh 2s + \cosh 2\lambda x )^{3/2}}.
\end{eqnarray}

The new field configuration (\ref{eq:sphaleron_proposal}), unlike the ansatz (\ref{eq:oscillon_proposal}), is entirely based on sphaleron d.o.f. The associated effective model can be computed analytically. However, due to the length of some of the expressions involved, it has been omitted here. As in the oscillon-based model, we consider the cases $s = 0.1$ and $s = 1.267$ with the same initial conditions for field theory and turning points. Once more, the initial conditions for $a(0),\, b(0)$ and $\lambda$ are determined by fitting the profile (\ref{eq:sphaleron_proposal}) with the field theory profile at the turning points. The values of the field configuration at the origin $\phi(0,t)$ in field theory and in the sphaleron-based effective model (\ref{eq:sphaleron_proposal}) are shown in Fig. \ref{fig:comparison_barrier_01} and in Fig. \ref{fig:comparison_barrier_1267}. The corresponding oscillation and modulation frequencies are presented in Table \ref{tab:oscillation} and in Table \ref{tab:modulation} respectively.

In light of the results obtained, the sphaleron-based model adapted to the oscillon (\ref{eq:sphaleron_proposal}) also determines the involved frequencies with high accuracy. The agreement is comparable to the one obtained through the oscillon-based model (\ref{eq:oscillon_proposal}). Similar results are obtained for other choices of the $s-$parameter.

It seems therefore that two d.o.f are enough to describe the oscillon evolution after a decay of a sphaleron without internal modes. The Derrick mode can be interpreted in this context as a kind of internal excitation of the oscillon itself that is responsible for the amplitude modulation.    

%============================================
%============================================
%============================================
%============================================
%============================================

\subsection{CCM in the well model}

\begin{figure}[!ht]
\centering
\begin{subfigure}[b]{\columnwidth}
  \centering
  \includegraphics[width=0.91\linewidth]{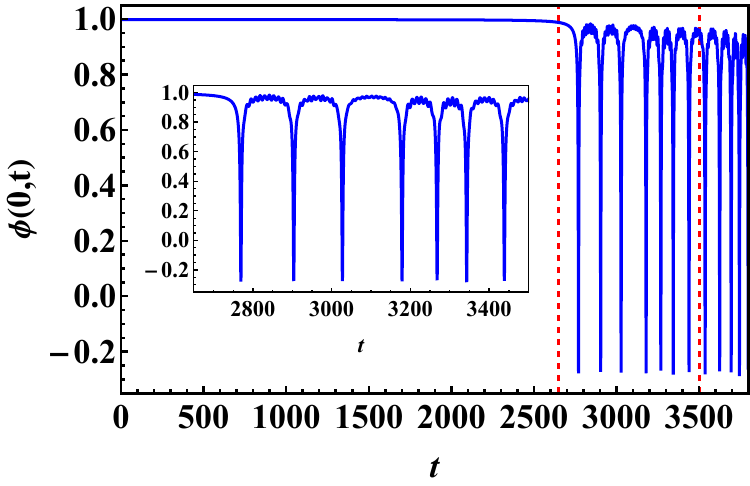}
  \caption{\small Well model: $s = 0.002$, $A = 0.01$\,.}
  \label{fig:decay_Well_Small_S}
\end{subfigure}

\vspace{0.3cm}

\begin{subfigure}[b]{\columnwidth}
  \centering
  \includegraphics[width=0.94\linewidth]{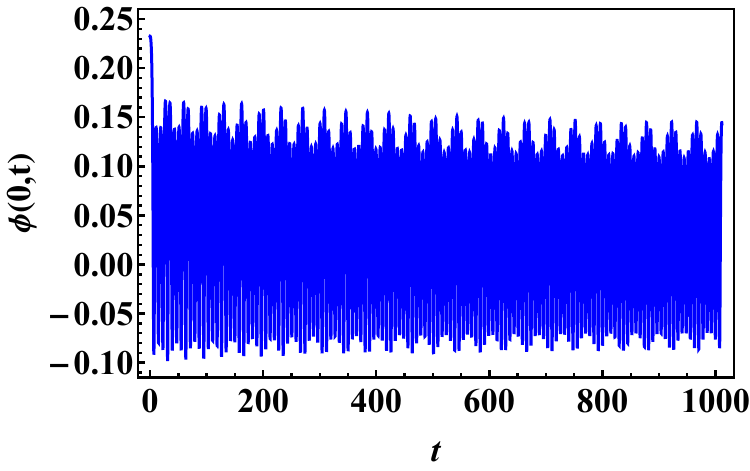}
  \caption{\small Well model: $s = 1.0$, $A = 0.01$\,.}
  \label{fig:decay_Well_Big_S}
\end{subfigure}
\caption{\small \justifying Upper panel: intermediate state of bounces for $s=0.002$ in the well model. The spectrum of linear perturbations contains $N=5$ even parity mode. The red dashed lines indicate the range shown in the inset. Lower panel: 
Decay into an oscillon for $s=1.0$ in the well model. The spectrum of linear perturbations contains $N=1$ even parity mode.}
%\label{fig:decay_Well_Big_S}
\end{figure}

The sphaleron decay is more intricate in the well model. When excited with small amplitudes of the negative mode and for small values of $s$ (see Fig. \ref{fig:decay_Well_Small_S}), there is an intermediate regime before the formation of an oscillon. After the first collapse the sphaleron is able to bounce a certain number of times. With each bounce, part of the energy is released until it settles into an oscillon state (this effect is also visible in Fig. \ref{fig:decay_well_large} for a range of amplitudes). The mechanism behind these bounces is intimately linked to the presence of the internal modes. As the sphaleron starts to decay part of the energy is transferred to the internal modes. These modes are able to store the energy for certain time and transfer it back in form of kinetic energy to the constituent subkinks. 

As $s$ grows the number of internal modes decreases, and this storing mechanism is not available. This means that, for $s$ big enough, the sphaleron should decay into an oscillon state without going through these intermediate bounces (see Fig. \ref{fig:decay_Well_Big_S}). Even for very large $s$, there is always at least one positive bound mode in the spectrum, but it does not seem to be enough to produce appreciable differences with respect to the barrier model studied in the previous section. 

In this section we will focus on the large $s$ regime and we will try to describe again the oscillon dynamics from the perspective of an oscillon-based model and an effective model based on a sphaleron profile adapted to the oscillon profile. Some interesting features of the small $s$ regime will be discussed in the following section.

Regarding the oscillon-based model, it is obtained again by introducing (\ref{eq:oscillon_proposal}) into (\ref{eq:Action}) with the potential given by (\ref{potential_Well}). Hence, the metric is the same as in (\ref{eq:metric}), and the only change appears in the effective potential. The calculation is straightforward but the result is too long and not very illuminating, therefore we have decided not to include it here. 

With respect to the adapted sphaleron model, we approximate the profile of the oscillon by
\begin{equation}
\Phi = a \,\left(1 - 2\,g(\lambda x;s)\sinh s\,)\right),    
\end{equation}
with $g(x;s)$ given by (\ref{eq:g_function}), and by performing a small scale deformation $\lambda \rightarrow\lambda + \epsilon$ and promoting the amplitudes to time-dependent parameters we are left with
\begin{eqnarray}\label{eq:Spaleron_proposal_well}
\Phi(x,t) &=& a(t)\,\left(1 - 2\,g(\lambda\, x;s)\sinh s\right)\\ \nonumber
&+& b(t)\,\dfrac{4\,x \sinh(x\, \lambda / 2) \sinh s}{\left( 3 + \cosh 2s + 2 \cosh \lambda\,x\sinh^2 s \right)^{3/2}}\,.
\end{eqnarray}

To determine the initial conditions $a(0),\, b(0)$ and $\lambda$, we adopt the same strategy as in the barrier model. First, we evolve the initial condition (\ref{IC_decay}) for a given amplitude $A$. Then, we adjust the field theory profile to our field proposal at an instant of time $t_0$ with no kinetic energy and $\dot{a}(0) = \dot{b}(0) = 0$ is assumed.

We have consider the cases $s = 2.16$ and $s = 1.0$, where both sphalerons hold only one even parity positive bound mode but their sizes are quite different. The initial unstable amplitudes for field theory have been $A = 0.001$ and $A = 0.01$ for each case. The turning points has been selected at the instants $t_0 = 7051.58$ and $t_0 = 7009.29$.

\begin{table}[!ht]
\centering
\renewcommand{\arraystretch}{1.5}
\setlength{\tabcolsep}{12pt}
\begin{tabular}{|c|c|c|c|}
\hline
\hline
\textbf{$s$} & \textbf{$\omega_{osc}^{ET(O)}$} & \textbf{$\omega_{osc}^{ET(S)}$} & \textbf{$\omega_{osc}^{FT}$} \\
\hline
\hline
2.16 & 0.920 & 0.923 & 0.918 \\
\hline
1.0 & 0.889 & 0.894 & 0.886 \\
\hline
\end{tabular}
\caption{\small \justifying Oscillation frequencies for $s = 2.16$ and $s = 1.0$ in field theory and in the effective models for the well model. The upper index $O$ or $S$ denotes the effective model based on the oscillon-based model or on the adapted sphaleron respectively.}
\label{tab:oscillation_2}
\end{table}

\begin{table}[!ht]
\centering
\renewcommand{\arraystretch}{1.5}
\setlength{\tabcolsep}{12pt}
\begin{tabular}{|c|c|c|c|}
\hline
\hline
\textbf{$s$} & \textbf{$\omega_{mod}^{ET(O)}$} & \textbf{$\omega_{mod}^{ET(S)}$} & \textbf{$\omega_{mod}^{FT}$} \\
\hline
\hline
2.16 & 0.09 & 0.088 & 0.078 \\
\hline
1.0 & 0.117 & 0.112 & 0.108 \\
\hline
\end{tabular}
\caption{\small \justifying Modulation frequencies for $s = 2.16$ and $s = 1.0$ in field theory and in the effective models for the well model. The upper index $O$ or $S$ denotes the effective model based on the oscillon-based model or on the adapted sphaleron respectively.}
\label{tab:modulation_2}
\end{table}

\begin{figure*}[!ht]
    \begin{subfigure}[b]{0.331\textwidth}
        \centering
        \includegraphics[width=0.99\columnwidth]{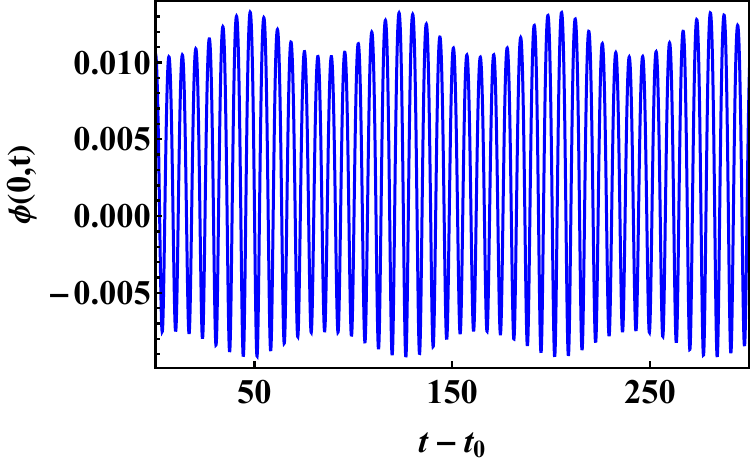}
        \caption{\small Field theory\,.}
        \label{Decay_Well_s216_FT}
    \end{subfigure}%
    \begin{subfigure}[b]{0.331\textwidth}
        \centering
        \includegraphics[width=0.99\columnwidth]{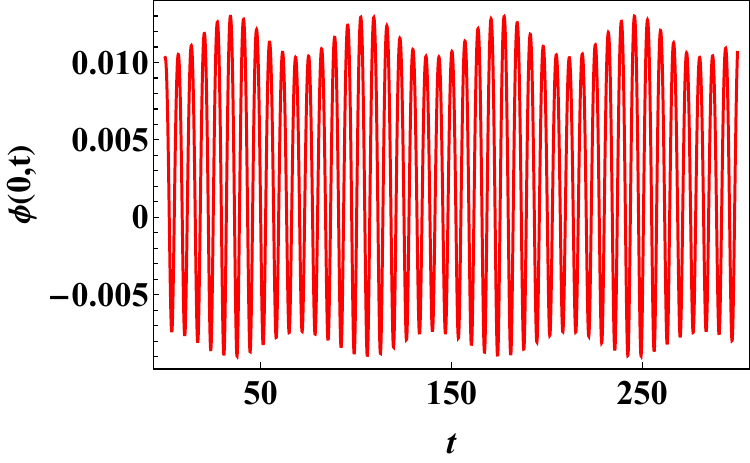}
        \caption{\small Oscillon-based model \,.}
        \label{LeastSquare_Well_216_Osc}
    \end{subfigure}
    \begin{subfigure}[b]{0.331\textwidth}
        \centering
        \includegraphics[width=0.99\columnwidth]{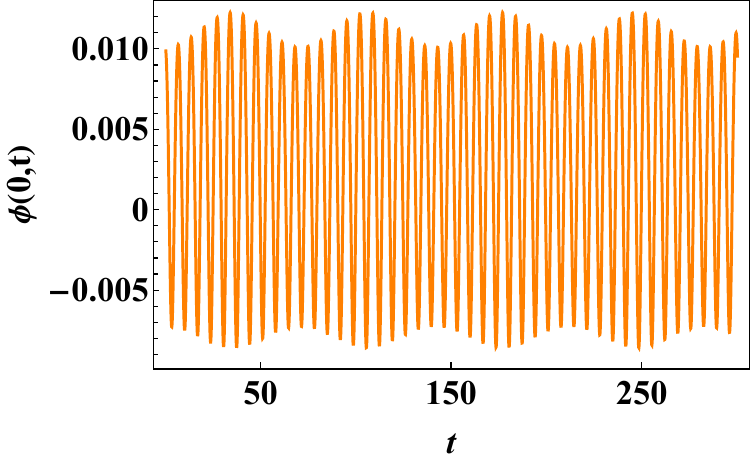}
        \caption{\small Sphaleron-based model\,.}
        \label{LeastSquare_Well_216_Sph}
    \end{subfigure}
    \caption{\small \justifying Comparison between field theory dynamics (left panel), and the least squares fitting in the oscillon-based model (middle panel) and in the sphaleron-based (right panel) for $s = 2.16$ in the well model.}
    \label{fig:comparison_well_216}
\end{figure*}

\begin{figure*}[!ht]
    \begin{subfigure}[b]{0.331\textwidth}
        \centering
        \includegraphics[width=0.99\columnwidth]{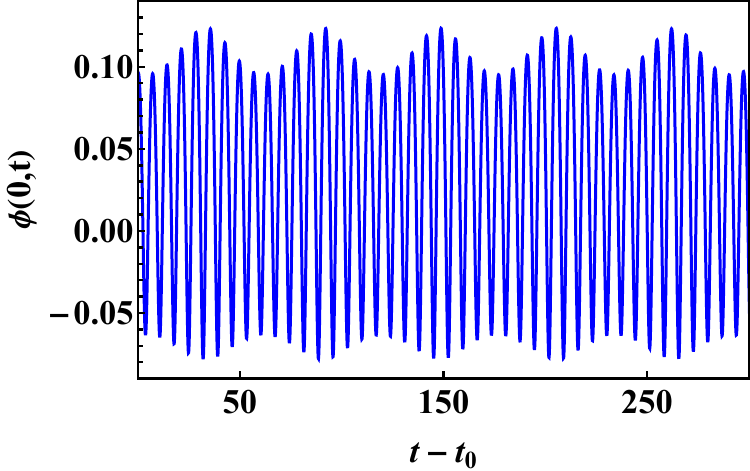}
        \caption{\small Field theory\,.}
        \label{Decay_Well_s1_FT}
    \end{subfigure}%
    \begin{subfigure}[b]{0.331\textwidth}
        \centering
        \includegraphics[width=0.99\columnwidth]{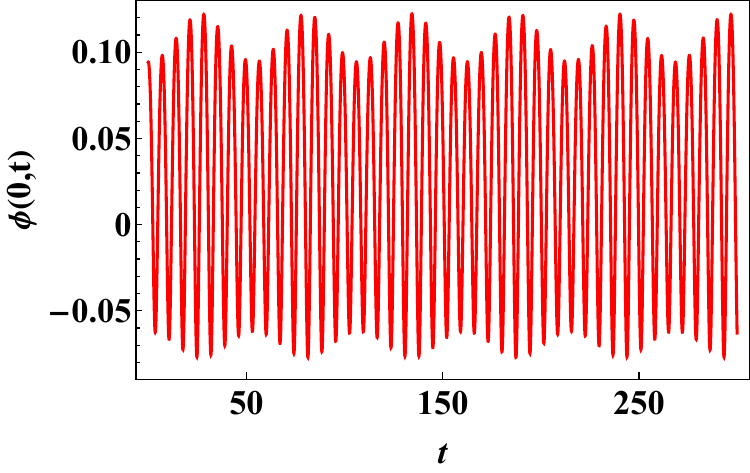}
        \caption{\small Oscillon-based model \,.}
        \label{LeastSquare_Well_1_Osc}
    \end{subfigure}
    \begin{subfigure}[b]{0.331\textwidth}
        \centering
        \includegraphics[width=0.99\columnwidth]{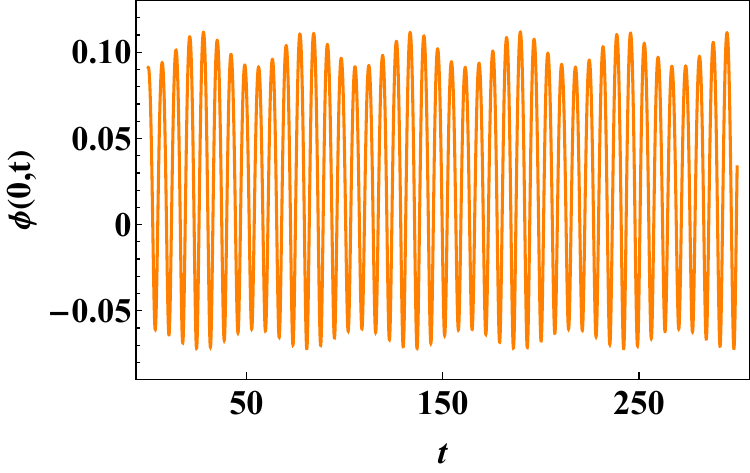}
        \caption{\small Sphaleron-based model\,.}
        \label{LeastSquare_Well_1_Sph}
    \end{subfigure}
    \caption{\small \justifying Comparison between field theory dynamics (left panel), and the least squares fitting in the oscillon-based model (middle panel) and in the sphaleron-based (right panel) for $s = 1.0$ in the well model.}
    \label{fig:comparison_well_1}
\end{figure*}

In Fig. \ref{fig:comparison_well_216} and Fig. \ref{fig:comparison_well_1} we illustrate the values of the field at the origin $\phi(0,t)$ in field theory and the result obtained from the effective models based on (\ref{eq:oscillon_proposal}) and (\ref{eq:Spaleron_proposal_well}) for $s = 2.16$ and $s = 1.0$ respectively. It is appreciable that these models are able to reproduce the amplitude modulation of the oscillon. The corresponding frequencies are collected in Table \ref{tab:oscillation_2} and Table \ref{tab:modulation_2}. It is again clearly visible that the oscillation frequency $\omega_{osc}$ is captured with high precision by the effective models, and only a slight deviation is appreciated in the modulation frequency.

Finally, we would like to mention that the presence of shapes modes in this model could allow us to introduce an ansatz for the field configuration based on the sphaleron profile along with the positive bound modes as follows
\begin{eqnarray}
\hspace{-0.3cm}
\Phi(x,t) &=& \phi_{S}(x;s) \nonumber\\
&+& a(t)\eta_{-1}(x;s) + \sum_{i = 1}^{N} b_i(t)\eta_{i}(x;s).\label{eq:modes_well_ansatz}
\end{eqnarray}
Here $N$ accounts for the maximum number of shape modes for a given value of $s$, and all the internal modes are assumed normalized. This leads to the following Lagrangian
\begin{equation}\label{eq:Lag_well}
L[a,b_i] = \dfrac{1}{2}\dot{a}^2 + \dfrac{1}{2}\sum_{i = 1}^{N}\dot{b}_i^2  - V_{eff}(a,b_i).
\end{equation}
The orthogonality of the eigenfunctions (Sturm-Liouville problem) ensures that the metric on the moduli space is diagonal. The complete expression of $V_{eff}(a,b_i)$ is extremely lengthy and can only be determined numerically. Therefore, it has been omitted here as it does not provide significant insight.

Contrary to expectations, the tests performed with this proposal for the previous cases do not show significant agreement with the field theory simulations. This is related to the inability of (\ref{eq:modes_well_ansatz}) to reproduce the oscillon profile, specially for small values of $s$, where the initial sphaleron and therefore the shape modes are notably bigger than the subsequent oscillon. Although it may be some range of big $s$-values for which the accordance could be better, there is not mathematical or physical guarantee for this to happen.

%============================================
%============================================
%============================================
%============================================
%============================================

\subsection{KAK formation in the well model}

As it has been briefly commented in Section \ref{sec:decay}, an unexpected behavior appears when $s$ takes small values in the well model. In that limit, the sphaleron resembles a KAK pair in the original $\phi^6$ model. For sufficiently large amplitudes ($A \geq A_{crit}$) of the unstable mode, instead of collapsing, the sphaleron expands as two outgoing accelerating kinks.

\begin{figure}[!ht]
\centering
{
\includegraphics[width=0.93\columnwidth]{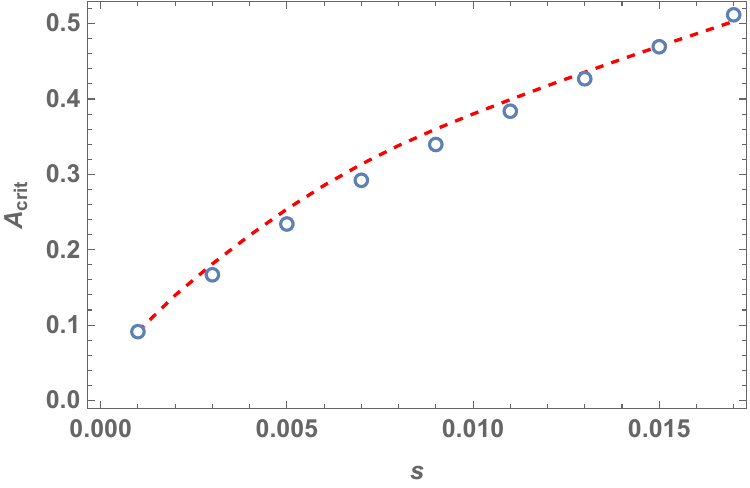}   
}
\caption{\small \justifying Dependence of the critical amplitude $A_{crit}$ on the model parameter $s$ in the well model. The dashed line corresponds to the theoretical prediction given by (\ref{eq:crit}) and the dots to the full numerical computation.}
\label{fig:critAmp}
\end{figure}

The dependence of this critical amplitude on the model parameter $s$ is illustrated in Fig. \ref{fig:critAmp}. It is important to remark that this phenomenon can be seen for considerably small values of the unstable amplitude $A$ as long as the parameter $s$ is small enough.

In order to explain this phenomenon, we have assumed an ansatz of the form (\ref{eq:modes_well_ansatz}) with $a(t) \sim \mathcal{O}\left( A \right)$ and $b_i(t) \sim \mathcal{O}\left( A^2 \right)$, where $A$ is the initial amplitude of the unstable mode. When introduced in the Lagrangian density and integrated over the space, the resulting effective Lagrangian up to $\mathcal{O}\left( A^4 \right)$ is
\begin{eqnarray}
L[a,b_i] &=& \dfrac{1}{2}\dot{a}^2 - \dfrac{1}{2}\omega_{-1}^2\,a^2 + \dfrac{1}{2}\sum_{i = 1}^{N}\dot{b}_i^2 - \dfrac{1}{2}\sum_{i = 1}^{N}\omega_{i}^2\,b_i^2\nonumber\\
&-& \alpha\, a^3 - \beta\, a^4 - \sum_{i = 1}^{N}\gamma_i\, a^2\, b_i\,. \label{eq:effective_Lagrangian}
\end{eqnarray}
The coefficients $\alpha, \, \beta$ and $\gamma_i$ are given explicitly by
\begin{eqnarray}
\alpha &=& 6 \int U_{s 6}^{'''}(\phi_S(x))\,\eta_{-1}^3(x)\, dx\,,\\
\beta&=& 24 \int U_{s 6}^{(4)}(\phi_S(x))\,\eta_{-1}^4(x) \, dx\,,\\
\gamma_i &=& 2 \int U_{s 6}^{''}(\phi_S(x))\,\eta_{-1}^2(x)\,\eta_{i}(x)\, dx\label{eq:gamma_i}\,.
\end{eqnarray} 
Solving the associated differential equation for $b_i$ at this order, we obtain
\begin{equation}\label{eq:mod_uns}
b_i(t) = A^2\,\dfrac{\gamma_i}{\omega_i^2}(\cos(\omega_i t) - 1),
\end{equation}
for the initial conditions
\begin{eqnarray}
a(0) = A,\hspace{0.6cm} b_i(0) = 0.
\end{eqnarray}
From (\ref{eq:gamma_i}) it can be seen that $\gamma_i$ only contributes for even indexes $i$, that is, only the shape modes with the same symmetry as the unstable mode are excited at lowest order. This approximate solution captures with good agreement the initial excitation of the shape modes in field theory (see Appendix \ref{sec:Excitation_shape}). 

Therefore, the excitation of the unstable mode in the collapsing direction has two implications: on the one hand, the size of the sphaleron is reduced, and on the other hand, the massive modes are triggered. As a result of the shrinking, a static attractive force between the constituent subkinks appears. We will show that the excitation of the massive modes may cancel this attraction. Let us start with the following approximate field configuration
\begin{equation}\label{crit_ansatz}
 \phi=\phi_0(x,A_c)+A_{-1}\eta_{-1}+\sum_i A_i\phi_i^{(1)}+\sum_i A_i^2\phi_i^{(2)},    
\end{equation}
where $\phi_0(x,A_c)=\phi_S(x)+A_c\,\eta_{-1}(x)$, that is, the initial deformed sphaleron in the direction of the unstable mode. $\phi^{(1)}$ and $\phi^{(2)}$ stands for the first and second order corrections to internal modes. The idea is simple: solve the equation order by order in $A_i$ given by the ansatz (\ref{crit_ansatz}), and look for the value $A_c$ such that $A_{-1}$ is purely oscillatory. This leads to the stationary solution we are looking for. By substituting (\ref{crit_ansatz}) into (\ref{eq:well}) and projecting onto the unstable mode, we get at zeroth order in $A_i$ and first order in $A_1$
\begin{equation}\label{eq:Am1}
  \ddot{A}_{-1}+\omega_{-1}^2A_{-1}=-A_c\,\omega_{-1}^2.  
\end{equation}

The first order equation in $A_i$ corresponds to the Schrodinger equation of the modes, therefore
\begin{equation}\label{eq:modi}
    \phi_i^{(1)}=\eta_i(x)\cos(\omega_i t).
\end{equation}
At second order in $A_i$ we get
\begin{equation}\label{eq:second_mod}
    \square \phi_i^{(2)}+U''(\phi_S)\phi_i^{(2)}=-\frac{1}{2}U'''(\phi_S)\left(\sum_i \phi_i^{(1)} \right)^2.
\end{equation}
Taking into account (\ref{eq:Am1}) and (\ref{eq:second_mod}) we finally get
\begin{eqnarray}
      \ddot{A}_{-1}&+&\omega_{-1}^2A_{-1}=-A_c\,\omega_{-1}^2 \nonumber \\ 
      && \hspace{-1.1cm} - \frac{1}{2}\int U'''(\phi_S)\,\eta_{-1}(x)\left(\sum_i A_i \phi_i^{(1)}\right)^2\,dx. \label{eq:tot_neg}
\end{eqnarray}

The inhomogeneous part of (\ref{eq:tot_neg}) consist of constants and oscillatory terms of the form $\cos(\omega_i t)\cos(\omega_j t)$. The latter produce oscillations of $A_{-1}$ which do not lead to exponential growing. The vanishing condition for the constant terms give precisely the critical value of the amplitude. Taking into account (\ref{eq:modi}) and (\ref{eq:mod_uns}) and disregarding the oscillatory part we obtain
\begin{equation}\label{eq:Aminus1}
      \ddot{A}_{-1}+\omega_{-1}^2A_{-1}=P(A_c),
\end{equation}
where
\begin{eqnarray}
    P(A_c)=-A_c\,\omega_{-1}^2-\frac{1}{4} A_c^4\sum_i \frac{\gamma_i^2\delta_i}{\omega_i^4}\,,\\
    \delta_i=\int U'''(\phi_S)\,\eta_{-1}(x)\,\eta_i(x)^2\,dx\,.
\end{eqnarray}

The amplitude $A_c$ which cancels the right hand side of (\ref{eq:Aminus1}) is precisely the value that leads to a purely oscillatory solution. Therefore, imposing $P(A_c)=0$ we get
\begin{equation}\label{eq:crit}
    A_c=-\left(\frac{4\,\omega_{-1}^2}{\sum\limits_i \frac{\gamma_i^2 \delta_i}{\omega_i^4}}\right)^{1/3}\,.
\end{equation}

In Fig. \ref{fig:critAmp} we compare the theoretical prediction (\ref{eq:crit}) with the full numerical computation. For the numerical calculation, we vary the amplitude $A$ of the unstable mode until we achieve a stationary solution. The analytical expression (\ref{eq:crit}) requires the calculation of the sphaleron modes and eigenvalues for each value of $s$. There is a good agreement between both computations. For large values of $s$ (small sphalerons) the stationary solution requires large amplitudes of the unstable mode departing from the linear regime. 
A similar phenomenon, called thick spectral wall, was also observed in KAK pairs in near-BPS theories \cite{Adam:2019uat}.  

This phenomenon can be seen as a mechanism for stabilizing unstable solutions. As we have seen, the main ingredient responsible for this mechanism is the excitation of internal modes capable of compensating the instability.

Although we have studied this phenomenon for a couple of particular models built \textit{ad hoc} for this purpose, we want to emphasize that this property seems to be completely generic. It is expected that any model with unstable sphaleron-type solutions can be stabilized, as long as the model hosts internal modes.

%============================================
%============================================
%============================================
%============================================
%============================================

\section{Summary and conclusions}\label{sec:summary}

In this paper we have studied the decay of unstable sphaleron-type solutions in two deformations of the $\phi^6$ model. The reason for the study of these two models is threefold: first they both describe KAK configurations of the $\phi^6$ model in a certain limit. Second, both contain analytical sphaleron-type solutions and third, despite their similarity, only one of them possesses positive internal modes. This feature results in crucial differences regarding the dynamics of the decay. 

In the first case, the $\phi^6$ barrier model, the sphaleron does not host positive internal modes. Excitations of the sphaleron in the unstable direction of contraction simply causes it to collapse. Excitations in the opposite direction produces accelerated expansion of the sphaleron. After collapsing, the sphaleron decays into an oscillon, which radiates very slowly. We have studied the oscillon resulting from the decay from various points of view. First we use as effective d.o.f. the amplitude of the leading term in a small amplitude expansion \cite{Navarro,Fodor:2008es} and its Derrick mode. In the second proposal, we use a profile adapted to the oscillon size based on the sphaleron itself. Typically, the oscillon possesses a modulated amplitude characterized by two frequencies, which is also observed in the effective models. 

In the second case, the $\phi^6$ well model, the sphaleron hosts positive internal modes, in increasing number as the $s$ parameter that measures the asymmetry between the true vacua and the false vacuum becomes smaller. For large $s$, as in the barrier case, after the collapse the sphaleron decays into an oscillon with a modulated amplitude.  For small $s$ the picture is quite different. For small amplitude excitations of the unstable mode the sphaleron collapses and bounces. The mechanism behind this phenomenon is intimately related to the presence of positive internal modes. During the collapse, the modes are able to store energy which is later transferred back to the constituent subkinks (the sphaleron walls) in the form of kinetic energy. Even more interestingly, as the sphaleron gets squeezed in the unstable direction, there is a critical value for which a stationary solution is formed. The excitation of the internal modes compensates the attraction in the unstable direction leading to an oscillatory sphaleron. Above this critical value the sphaleron solution expands. It is also interesting to note that this mechanism is only activated when the sphaleron contracts, since the internal modes always push the constituent subkinks outward.

From a more general perspective, although we have carried out a particular analysis in a model in $1+1$ dimensions built for this purpose, the mechanisms described here can exist in any model with unstable sphaleron-type solutions, as long as they have internal modes. We would like to emphasize that, even the formulas that determine the critical amplitude are rather generic and can be applied to any model hosting sphaleron-like solutions. Of course, achieving the stationary solution for a generic perturbation requires fine-tuning the excitation of the unstable mode. However, we have shown that determining in which direction the sphaleron will decay depends crucially on the critical amplitude of the negative mode, which, for a generic perturbation, will depend on the overlap of the latter with the mode. Therefore, the role  of this mechanism on the dynamics of unstable objects can be of great interest.

\section*{Acknowledgments}

We would like to thank A. Wereszczynski for his comments and fruitful discussions. S. N. O. also thanks C. Naya, T. Romanczukiewicz and K. Slawinska for discussions during his stay at Jagiellonian University.
This research was supported in part by Spanish MCIN with funding from European Union NextGenerationEU (PRTRC17.I1) and Consejeria de Educacion from JCyL through QCAYLE project, as well as MCIN project PID2020-113406GB-I00. S. N. O. acknowledge financial support from the European Social Fund, the Operational Programme of Junta de
Castilla y Leon and the regional Ministry of Education.

\begin{figure*}[!ht]
    \begin{subfigure}[b]{0.331\textwidth}
        \centering
        \includegraphics[width=1\columnwidth]{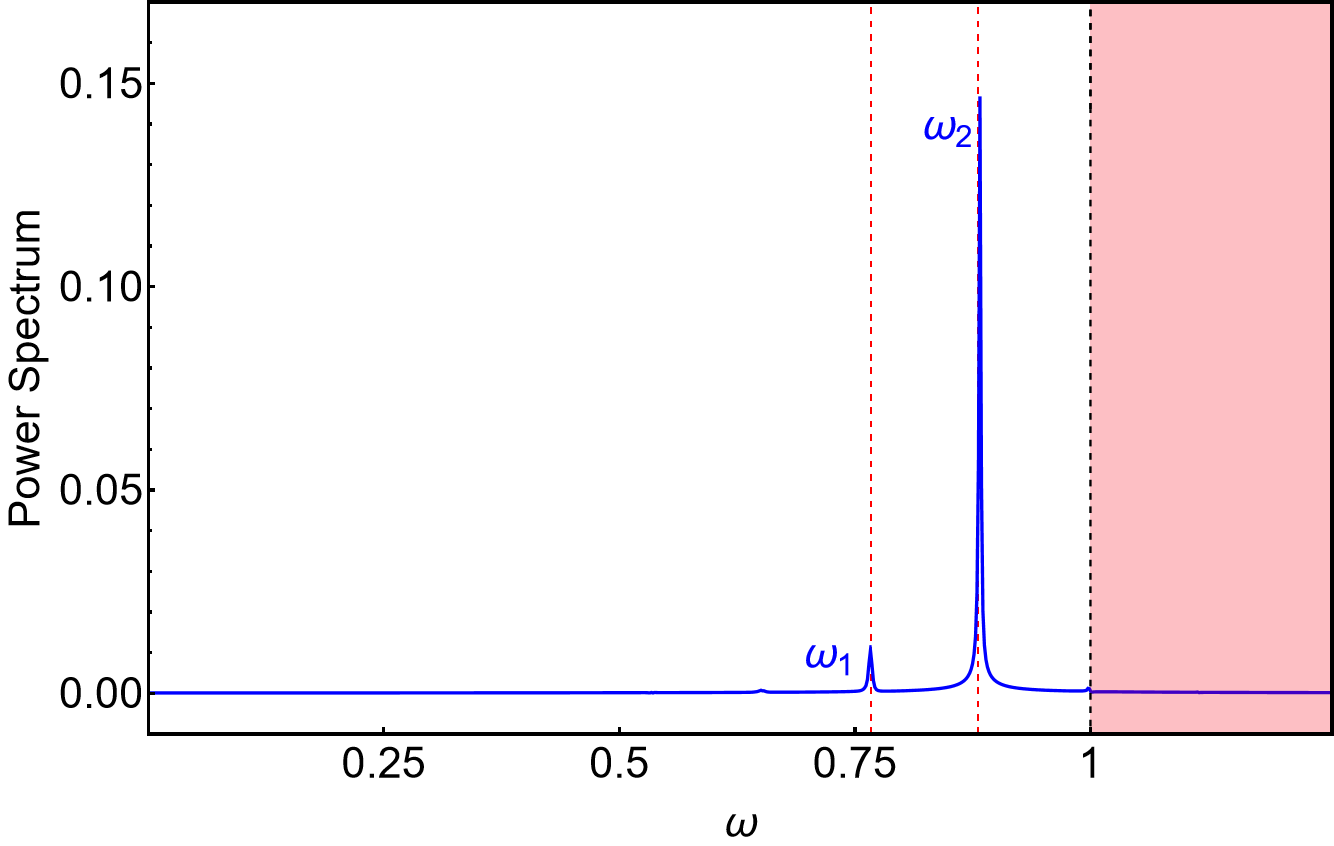}
        \caption{\small Field theory\,.}
        \label{FFT_FT}
    \end{subfigure}%
    \begin{subfigure}[b]{0.331\textwidth}
        \centering
        \includegraphics[width=1\columnwidth]{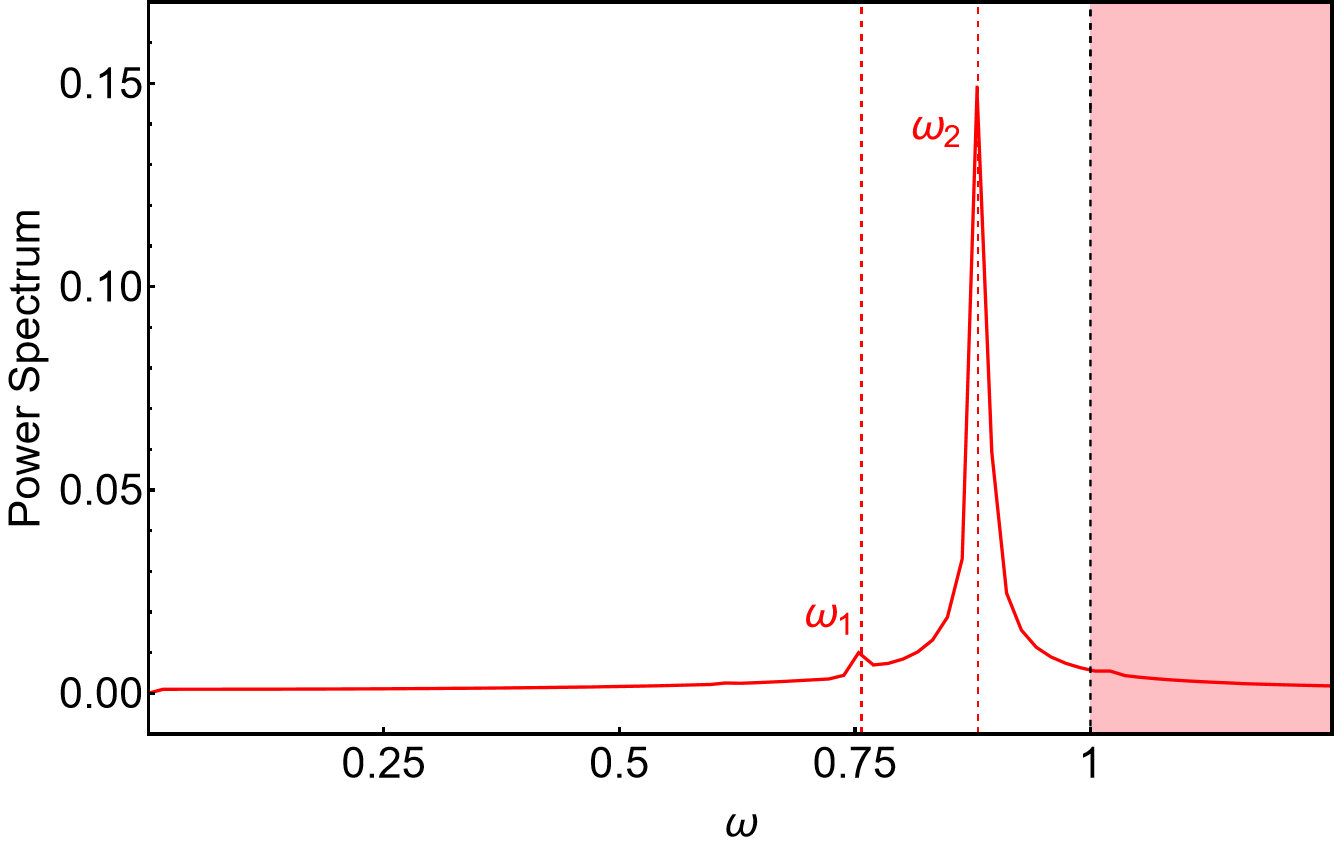}
        \caption{\small Oscillon-based effective model\,.}
        \label{FFT_Osc}
    \end{subfigure}
    \begin{subfigure}[b]{0.331\textwidth}
        \centering
        \includegraphics[width=1\columnwidth]{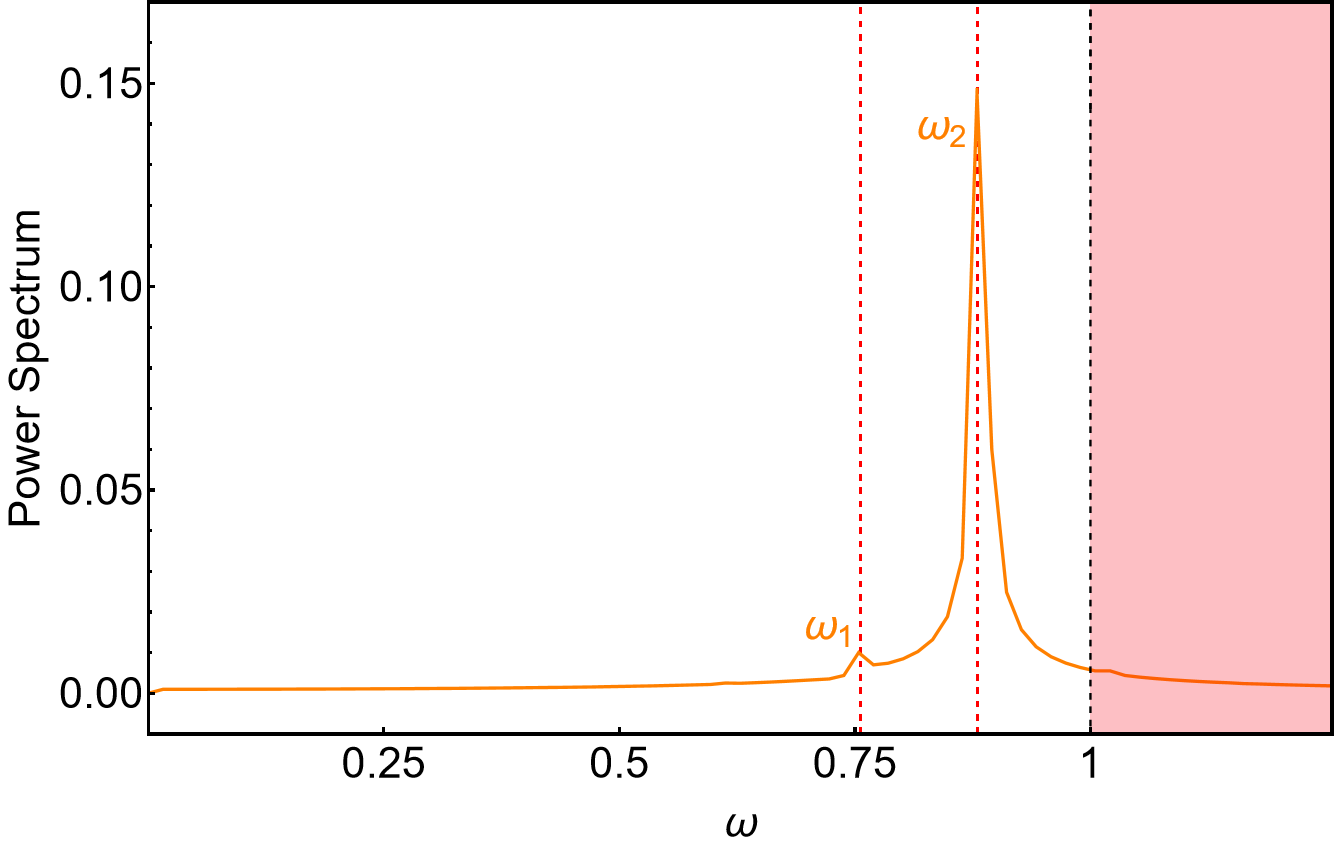}
        \caption{\small Sphaleron-based effective model\,.}
        \label{FFT_Sph}
    \end{subfigure}
    \caption{\small \justifying Fourier transform of the field values at the origin $\phi(0,t)$ for the barrier model. The shaded region represent the continuum, and the vertical dashed lines the main frequencies involved in the oscillon evolution for $s = 0.1$.}
    \label{fig:FFT_01}
\end{figure*}

\begin{figure*}[!ht]
    \begin{subfigure}[b]{0.331\textwidth}
        \centering
        \includegraphics[width=1\columnwidth]{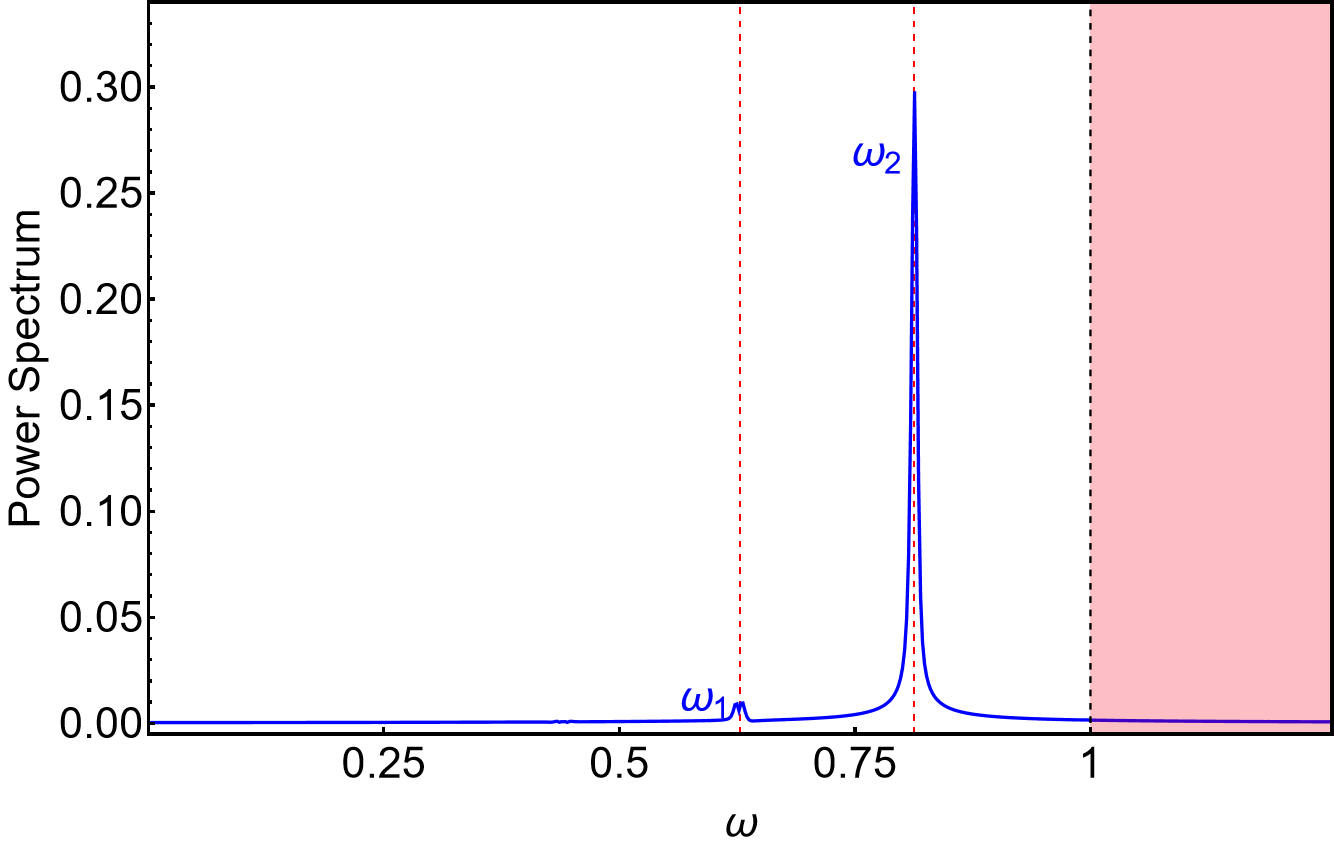}
        \caption{\small Field theory\,.}
        \label{FFT_FT_2}
    \end{subfigure}%
    \begin{subfigure}[b]{0.331\textwidth}
        \centering
        \includegraphics[width=1\columnwidth]{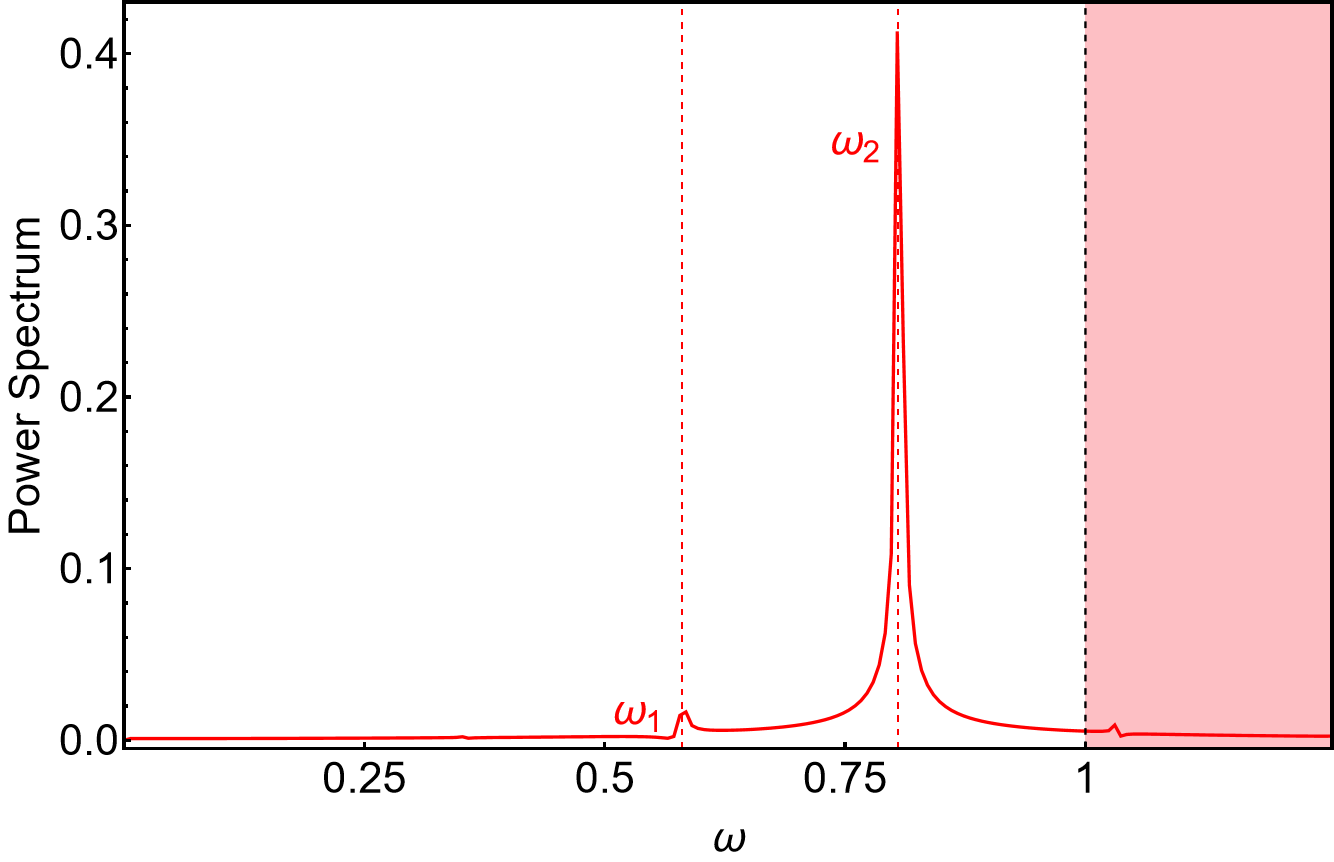}
        \caption{\small Oscillon-based effective model\,.}
        \label{FFT_Osc_2}
    \end{subfigure}
    \begin{subfigure}[b]{0.331\textwidth}
        \centering
        \includegraphics[width=1\columnwidth]{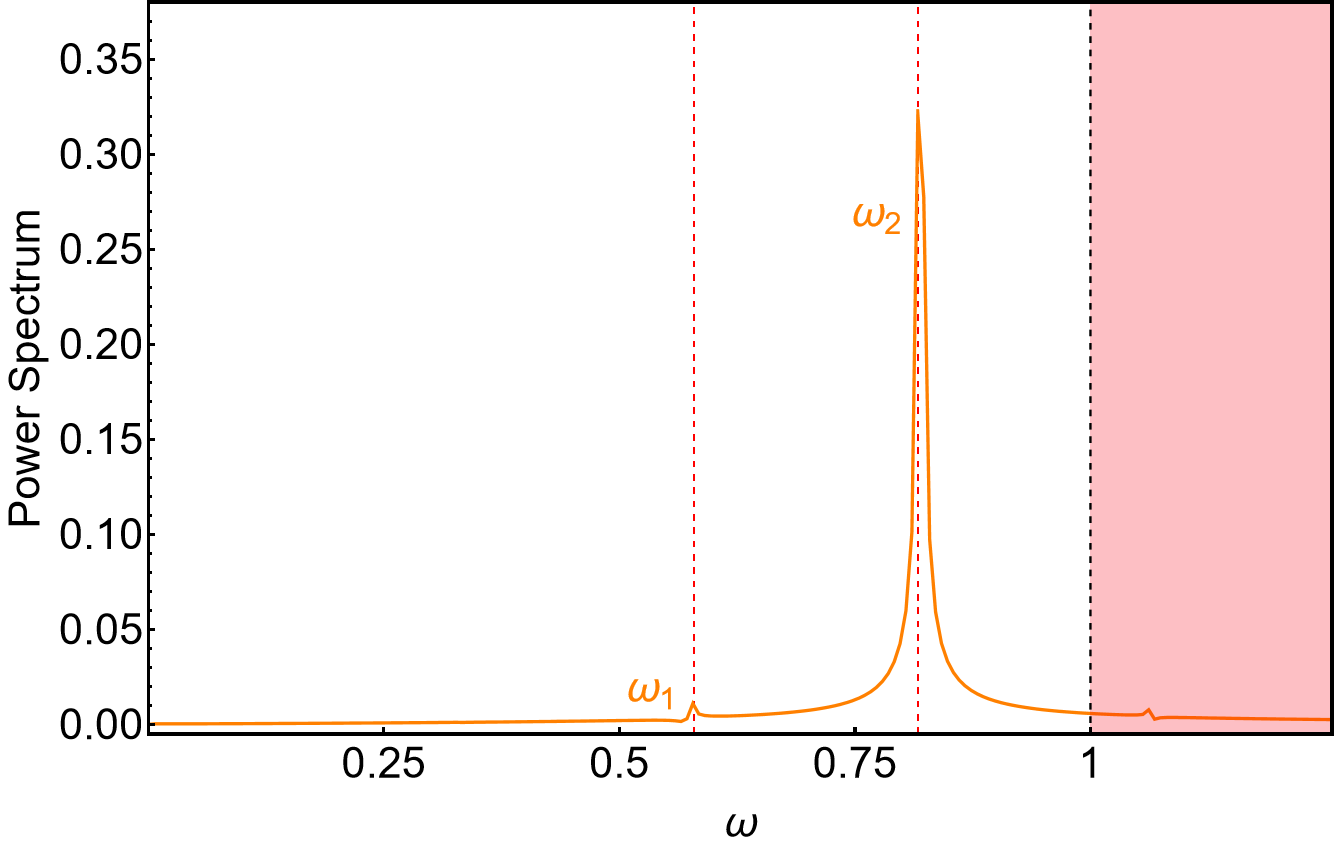}
        \caption{\small Sphaleron-based effective model\,.}
        \label{FFT_Sph_2}
    \end{subfigure}
    \caption{\small \justifying Fourier transform of the field values at the origin $\phi(0,t)$ for the barrier model. The shaded region represent the continuum, and the vertical dashed lines the main frequencies involved in the oscillon evolution for $s = 1.267$.}
    \label{fig:FFT_1267}
\end{figure*}

\begin{figure*}[!ht]
    \begin{subfigure}[b]{0.331\textwidth}
        \centering
        \includegraphics[width=1\columnwidth]{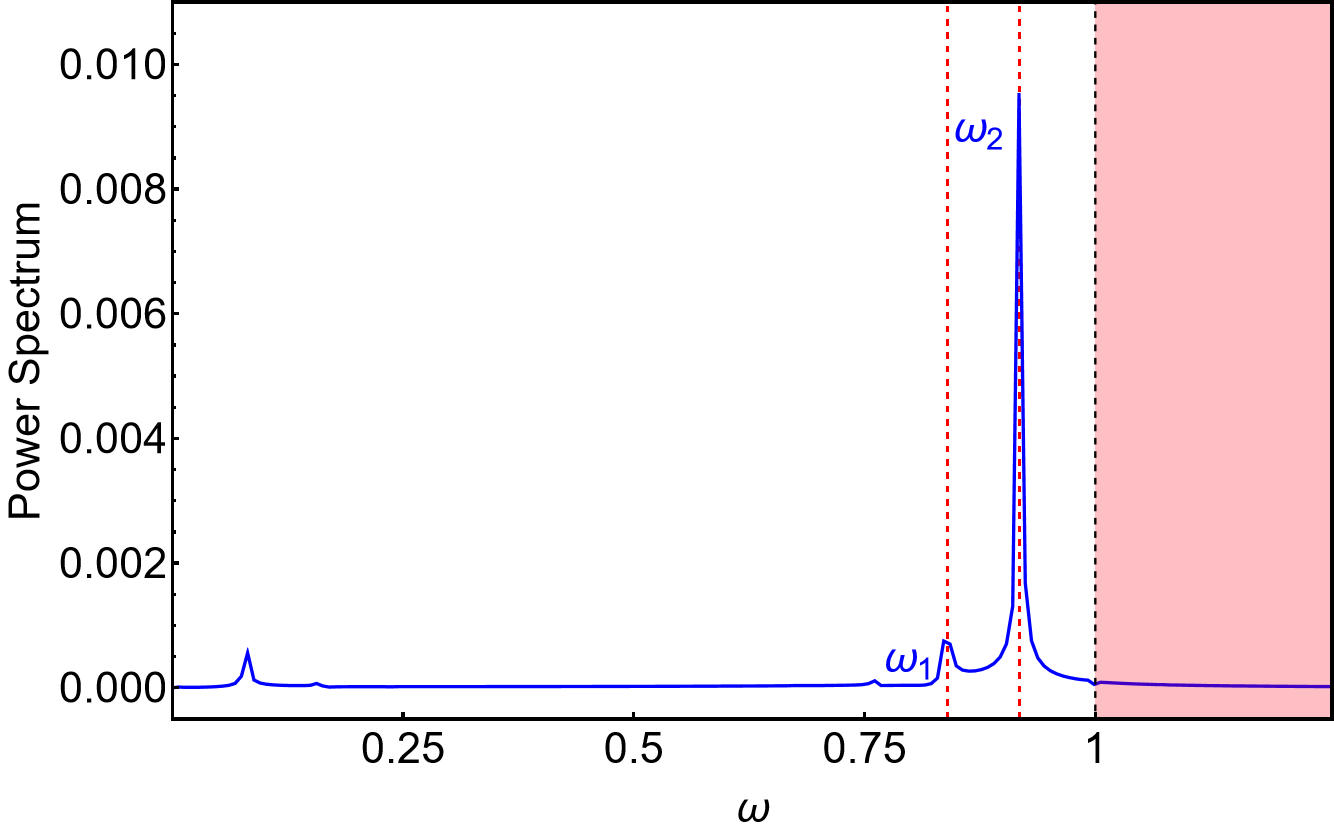}
        \caption{\small Field theory\,.}
        \label{FFT_FT_Well}
    \end{subfigure}%
    \begin{subfigure}[b]{0.331\textwidth}
        \centering
        \includegraphics[width=1\columnwidth]{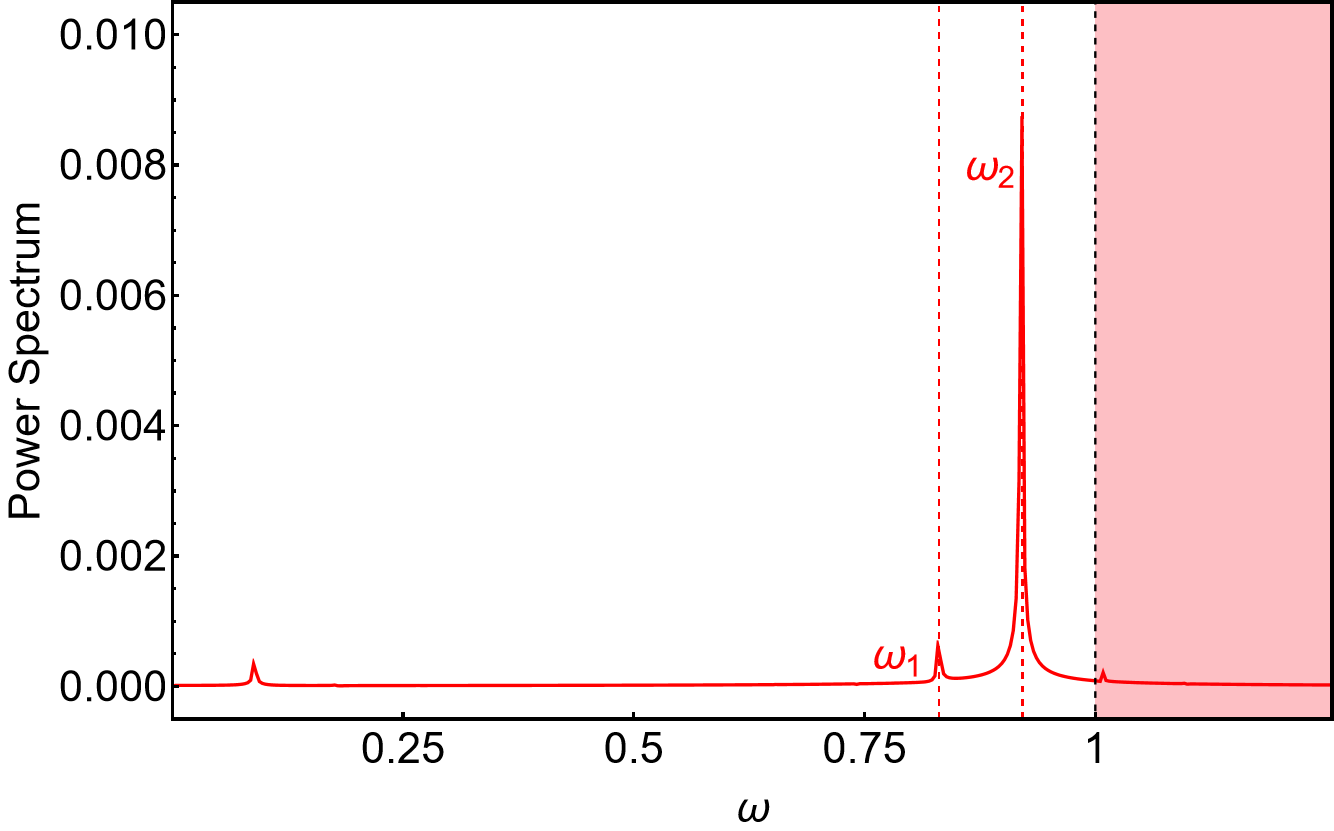}
        \caption{\small Oscillon-based effective model\,.}
        \label{FFT_Osc_Well}
    \end{subfigure}
    \begin{subfigure}[b]{0.331\textwidth}
        \centering
        \includegraphics[width=1\columnwidth]{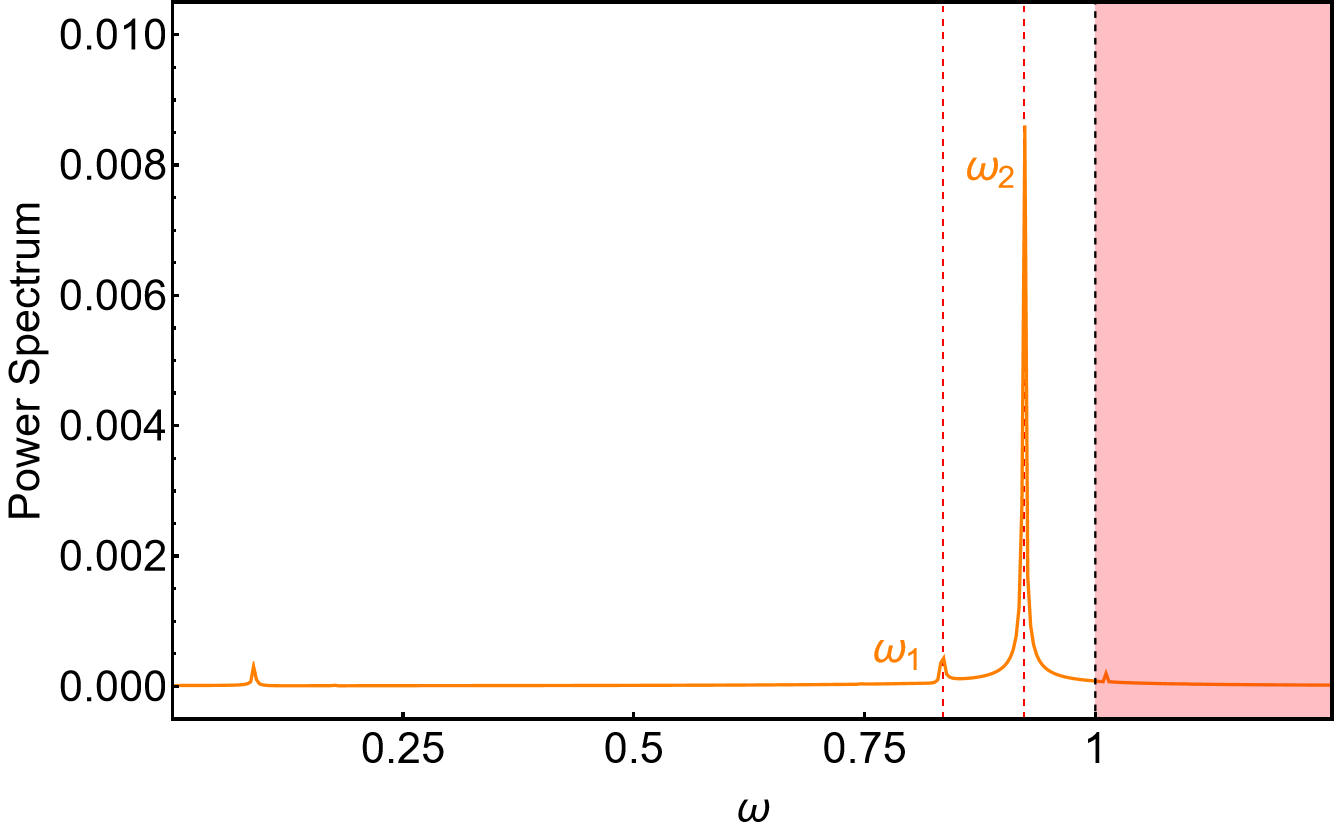}
        \caption{\small Sphaleron-based effective model\,.}
        \label{FFT_Sph_Well}
    \end{subfigure}
    \caption{\small \justifying Fourier transform of the field values at the origin $\phi(0,t)$ for the well model. The shaded region represent the continuum, and the vertical dashed lines the main frequencies involved in the oscillon evolution for $s = 2.16$.}
    \label{fig:FFT_216}
\end{figure*}

\begin{figure*}[!ht]
    \begin{subfigure}[b]{0.331\textwidth}
        \centering
        \includegraphics[width=1\columnwidth]{FFT_FT_well_s216_A0001_New.png}
        \caption{\small Field theory\,.}
        \label{FFT_FT_Well_2}
    \end{subfigure}%
    \begin{subfigure}[b]{0.331\textwidth}
        \centering
        \includegraphics[width=1\columnwidth]{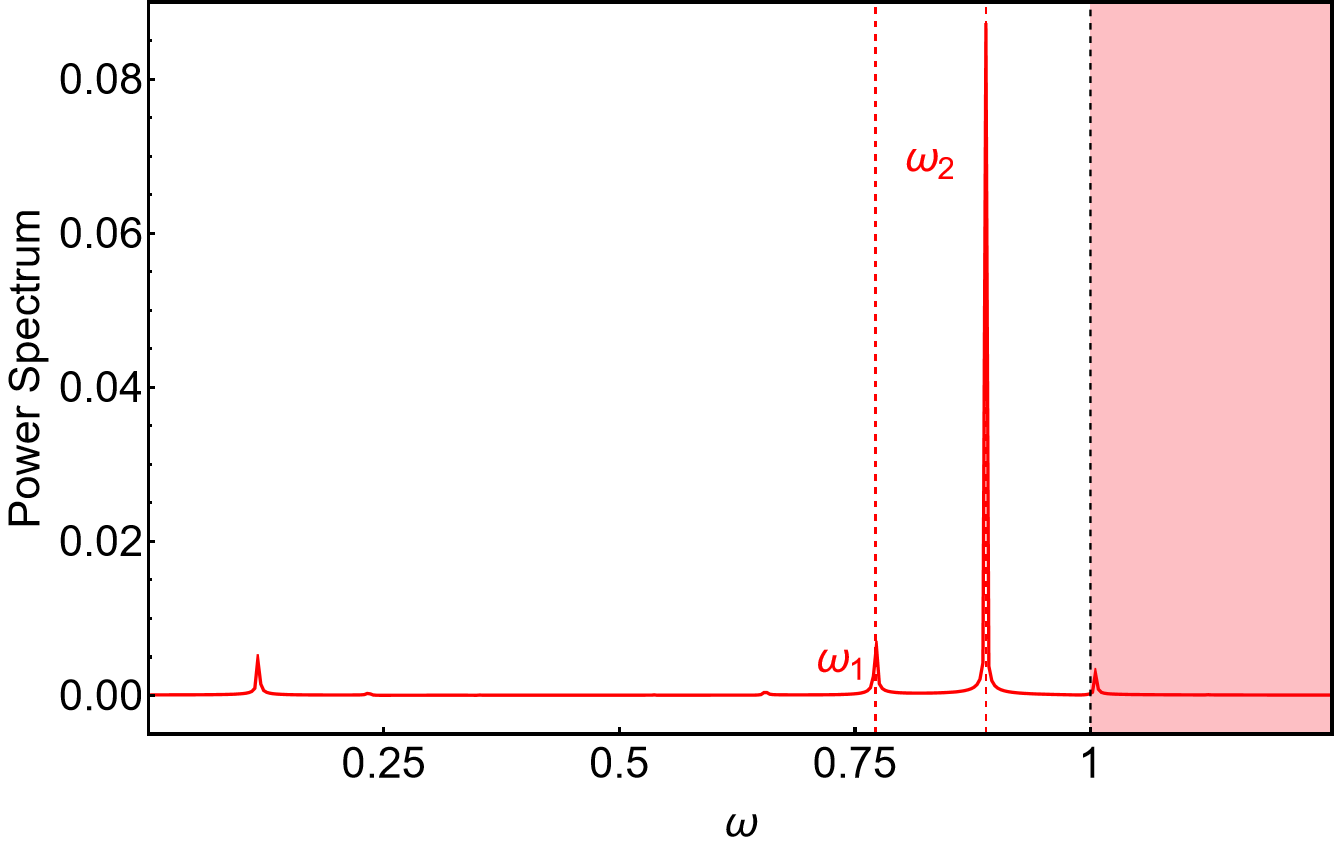}
        \caption{\small Oscillon-based effective model\,.}
        \label{FFT_Osc_Well_2}
    \end{subfigure}
    \begin{subfigure}[b]{0.331\textwidth}
        \centering
        \includegraphics[width=1\columnwidth]{FFT_Sph_well_s216_A0001_New.png}
        \caption{\small Sphaleron-based effective model\,.}
        \label{FFT_Sph_Well_2}
    \end{subfigure}
    \caption{\small \justifying Fourier transform of the field values at the origin $\phi(0,t)$ for the well model. The shaded region represent the continuum, and the vertical dashed lines the main frequencies involved in the oscillon evolution for $s = 1.0$.}
    \label{fig:FFT_1}
\end{figure*}

%============================================
%============================================
%============================================
%============================================
%============================================

\appendix

\section{Oscillation and modulation frequencies}\label{sec:FFT}

In the main part of this paper, it has been studied the frequencies involved in the oscilon dynamics after its formation from the decay of an initially perturbed sphaleron. It has been seen that the oscillon presents a double quasi-periodic structure, related to two main frequencies: the oscillation frequency $\omega_{osc}$ and the modulation frequency $\omega_{mod}$. These frequencies have been determined through the Fourier transform of the field configuration at the origin $\phi(0,t)$ using the scipy.fft library from Python.

In Fig. \ref{fig:FFT_01} and \ref{fig:FFT_1267} we represent the Fourier transforms computed in the barrier model for $s = 0.1$ and $s = 1.267$ respectively, and the Fourier transforms in the well model for $s = 2.16$ and $s = 1.0$ are depicted in Fig. \ref{fig:FFT_216} and  Fig. \ref{fig:FFT_1}. We have denoted the main frequencies by $\omega_1$ and $\omega_2$, where $\omega_{osc} = \omega_2$ and $\omega_{mod} = \omega_{2} - \omega_{1}$. The corresponding frequencies are collected in Table \ref{tab:oscillation}, Table \ref{tab:modulation}, Table \ref{tab:oscillation_2} and Table \ref{tab:modulation_2}.

%============================================
%============================================
%============================================
%============================================
%============================================

%============================================

\begin{figure*}[!ht]
  \centering
  \begin{minipage}[t]{0.45\textwidth}
    \centering
    \begin{subfigure}{\columnwidth}
      \includegraphics[width=1\columnwidth]{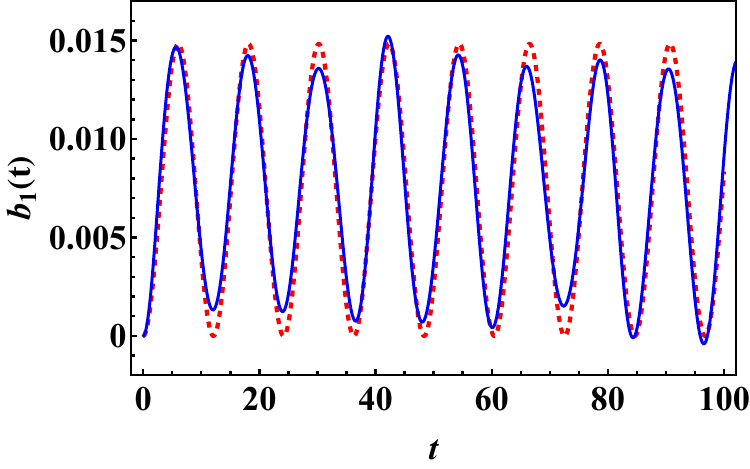}
      \caption{First bound mode.}
      \label{fig:b1}
    \end{subfigure}
    \vskip\baselineskip
    \begin{subfigure}{\columnwidth}
      \includegraphics[width=1\columnwidth]{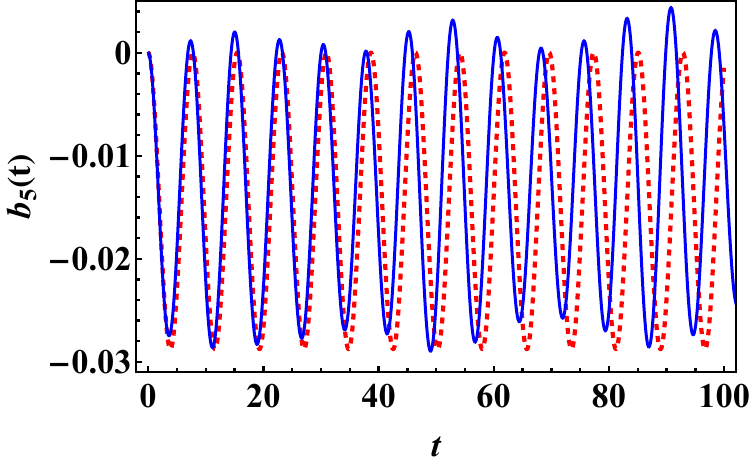}
      \caption{Fifth bound mode.}
      \label{fig:b5}
    \end{subfigure}
  \end{minipage}
  \hfill
  \begin{minipage}[t]{0.45\textwidth}
    \centering
    \begin{subfigure}{\columnwidth}
      \includegraphics[width=1\columnwidth]{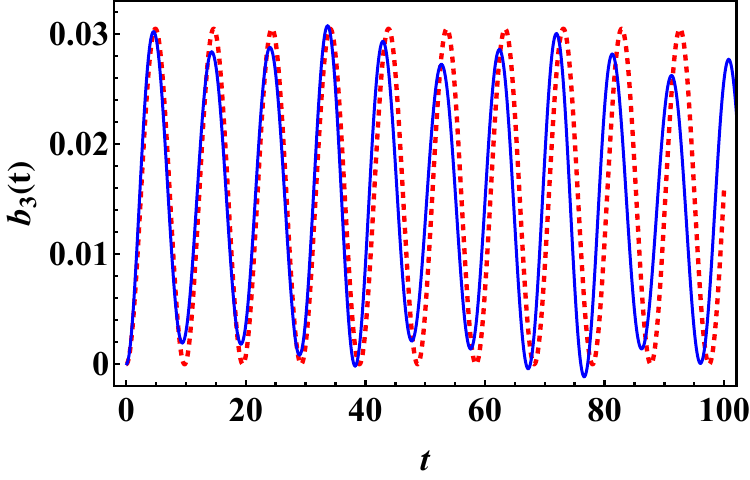}
      \caption{Third bound mode.}
      \label{fig:b3}
    \end{subfigure}
    \vskip\baselineskip
    \begin{subfigure}{\columnwidth}
      \includegraphics[width=1\columnwidth]{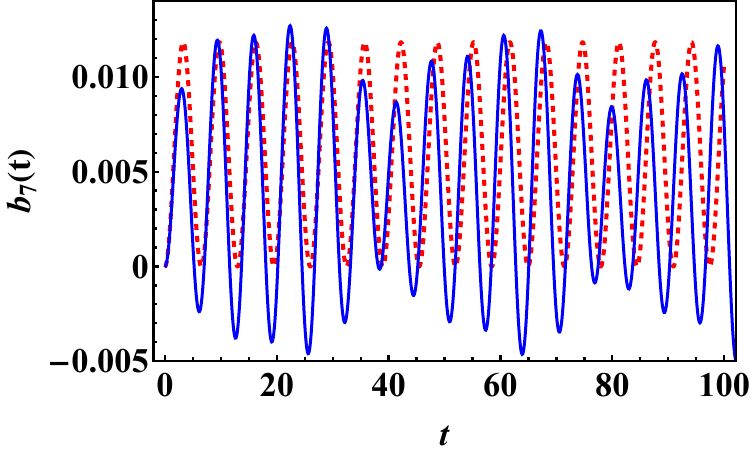}
      \caption{Seventh bound mode.}
      \label{fig:b7}
    \end{subfigure}
  \end{minipage}
  
  \caption{\small \justifying Excitation of the shape modes by their coupling to the unstable mode. The solid line represents the field theory dynamics (\ref{eq:projection}) and the dashed line the approximate analytical expression (\ref{eq:excitation_bi}).}
  \label{fig:excitations}
\end{figure*}

%============================================

\section{Excitation of the shape modes in the well model}\label{sec:Excitation_shape}

When the unstable mode of the sphaleron in the well model is initially excited, the nonlinear terms couple it to the available shape modes. This coupling can be determined at lowest order by approximating the field configuration by (\ref{eq:modes_well_ansatz}), with $a(t) \sim \mathcal{O}\left( A \right)$ and $b_i(t) \sim \mathcal{O}\left( A^2 \right)$, and where $A$ the initial amplitude of the unstable mode, obtaining the effective Lagrangian (\ref{eq:effective_Lagrangian}).

Solving the associated equations of motion for $b_i$, it has been obtained the following approximate solution 
\begin{equation}\label{eq:excitation_bi}
b_i(t) = A^2\,\dfrac{\gamma_i}{\omega_i^2}(\cos(\omega_i t) - 1),
\end{equation}
for the initial conditions
\begin{eqnarray}
a(0) = A,\hspace{0.6cm} b_i(0) = 0.
\end{eqnarray}
In order to verify the accuracy of (\ref{eq:excitation_bi}), it has been projected the profile $\phi(x,t)$ from field theory onto the shapes modes as follows
\begin{equation}\label{eq:projection}
b_{i}^{FT}(t) = \int \left(\phi(x,t) - \eta_{-1}(x;s) \right)\,\eta_{i}(x;s)\,dx\,.
\end{equation}
In Fig. \ref{fig:excitations} we compare the excitation of the modes with even symmetry in field theory with the approximate expression (\ref{eq:excitation_bi}) for $s = 0.005$ and $A = 0.26$.
There one can see initially a great agreement in the amplitude and in the oscillation frequency. The deviations present at longer times is expected to be related to higher order contributions and to the squeezing of the sphaleron, which changes its internal structure.

%============================================

\section{Details of the numerical simulations}\label{sec:NumericalAnalysis}

In this Appendix we summarize some relevant details of the numerical method that has been used during the simulations in field theory.

The equation of motion (\ref{eq:barrier}) or (\ref{eq:well}) has been discretized using an explicit second order finite difference scheme, and the simulations have been performed on the interval $- L < x < L$, with $L = 100$. Regarding the boundary conditions, Mur's conditions on both edges of the simulation box have been imposed
\begin{eqnarray}
\big(\partial_t \phi + \partial_x \phi\big)\Big|_{x = L,t} &=& 0\,,\\
\big(\partial_t \phi - \partial_x \phi\big)\Big|_{x = - L,t} &=& 0\,,
\end{eqnarray}
in order to prevent the radiation from reflecting back from
the boundaries of the system. Furthermore, a damping term of the form $\gamma(x) \phi_{t}$ has been introduced near the edges to absorb as much as possible any remaining amount of radiation. The explicit $x$-dependence of $\gamma(x)$ reads as \cite{Oxtoby}
\begin{equation}
\gamma(x) =
    \begin{cases}
      \left(\dfrac{x - (L - 10)}{10}\right)^4, & \text{if} \quad x \ge L - 10,\\
      \left(\dfrac{x + (L - 10)}{10}\right)^4, & \text{if} \quad x \leq - L + 10,\\
      \hspace{2.5cm} 0\,, & \text{otherwise.}
    \end{cases}
\end{equation}

%============================================
%============================================
%============================================
%============================================
%============================================

\bibliographystyle{apsrev4-2}
\bibliography{references}

\begin{thebibliography}{34}%
\makeatletter
\providecommand \@ifxundefined [1]{%
 \@ifx{#1\undefined}
}%
\providecommand \@ifnum [1]{%
 \ifnum #1\expandafter \@firstoftwo
 \else \expandafter \@secondoftwo
 \fi
}%
\providecommand \@ifx [1]{%
 \ifx #1\expandafter \@firstoftwo
 \else \expandafter \@secondoftwo
 \fi
}%
\providecommand \natexlab [1]{#1}%
\providecommand \enquote  [1]{``#1''}%
\providecommand \bibnamefont  [1]{#1}%
\providecommand \bibfnamefont [1]{#1}%
\providecommand \citenamefont [1]{#1}%
\providecommand \href@noop [0]{\@secondoftwo}%
\providecommand \href [0]{\begingroup \@sanitize@url \@href}%
\providecommand \@href[1]{\@@startlink{#1}\@@href}%
\providecommand \@@href[1]{\endgroup#1\@@endlink}%
\providecommand \@sanitize@url [0]{\catcode `\\12\catcode `\$12\catcode `\&12\catcode `\#12\catcode `\^12\catcode `\_12\catcode `\%12\relax}%
\providecommand \@@startlink[1]{}%
\providecommand \@@endlink[0]{}%
\providecommand \url  [0]{\begingroup\@sanitize@url \@url }%
\providecommand \@url [1]{\endgroup\@href {#1}{\urlprefix }}%
\providecommand \urlprefix  [0]{URL }%
\providecommand \Eprint [0]{\href }%
\providecommand \doibase [0]{https://doi.org/}%
\providecommand \selectlanguage [0]{\@gobble}%
\providecommand \bibinfo  [0]{\@secondoftwo}%
\providecommand \bibfield  [0]{\@secondoftwo}%
\providecommand \translation [1]{[#1]}%
\providecommand \BibitemOpen [0]{}%
\providecommand \bibitemStop [0]{}%
\providecommand \bibitemNoStop [0]{.\EOS\space}%
\providecommand \EOS [0]{\spacefactor3000\relax}%
\providecommand \BibitemShut  [1]{\csname bibitem#1\endcsname}%
\let\auto@bib@innerbib\@empty
%</preamble>
\bibitem [{\citenamefont {Manton}\ and\ \citenamefont {Sutcliffe}(2004)}]{Manton}%
  \BibitemOpen
  \bibfield  {author} {\bibinfo {author} {\bibfnamefont {N.}~\bibnamefont {Manton}}\ and\ \bibinfo {author} {\bibfnamefont {P.}~\bibnamefont {Sutcliffe}},\ }\href@noop {} {\emph {\bibinfo {title} {Topological Solitons}}}\ (\bibinfo  {publisher} {Cambridge University Press},\ \bibinfo {address} {Cambridge U.K.},\ \bibinfo {year} {2004})\BibitemShut {NoStop}%
\bibitem [{\citenamefont {Manton}(2019)}]{Manton2}%
  \BibitemOpen
  \bibfield  {author} {\bibinfo {author} {\bibfnamefont {N.}~\bibnamefont {Manton}},\ }\href@noop {} {\bibfield  {journal} {\bibinfo  {journal} {Phil. Trans. Roy. Soc. Lond. A.}\ }\textbf {\bibinfo {volume} {377}} (\bibinfo {year} {2019})}\BibitemShut {NoStop}%
\bibitem [{\citenamefont {Manton}\ and\ \citenamefont {Samols}(1988)}]{MSam}%
  \BibitemOpen
  \bibfield  {author} {\bibinfo {author} {\bibfnamefont {N.~S.}\ \bibnamefont {Manton}}\ and\ \bibinfo {author} {\bibfnamefont {T.~M.}\ \bibnamefont {Samols}},\ }\href@noop {} {\bibfield  {journal} {\bibinfo  {journal} {Phys. Lett. B}\ }\textbf {\bibinfo {volume} {207}},\ \bibinfo {pages} {179} (\bibinfo {year} {1988})}\BibitemShut {NoStop}%
\bibitem [{\citenamefont {Krusch}\ and\ \citenamefont {Sutcliffe}(2004)}]{KS}%
  \BibitemOpen
  \bibfield  {author} {\bibinfo {author} {\bibfnamefont {S.}~\bibnamefont {Krusch}}\ and\ \bibinfo {author} {\bibfnamefont {P.}~\bibnamefont {Sutcliffe}},\ }\href@noop {} {\bibfield  {journal} {\bibinfo  {journal} {J. Phys. A: Math. Theor.}\ }\textbf {\bibinfo {volume} {37}},\ \bibinfo {pages} {9037} (\bibinfo {year} {2004})}\BibitemShut {NoStop}%
\bibitem [{\citenamefont {Adam}\ \emph {et~al.}(2021)\citenamefont {Adam}, \citenamefont {Ciurla}, \citenamefont {Oles}, \citenamefont {Romanczukiewicz},\ and\ \citenamefont {Wereszczynski}}]{Adam}%
  \BibitemOpen
  \bibfield  {author} {\bibinfo {author} {\bibfnamefont {C.}~\bibnamefont {Adam}}, \bibinfo {author} {\bibfnamefont {D.}~\bibnamefont {Ciurla}}, \bibinfo {author} {\bibfnamefont {K.}~\bibnamefont {Oles}}, \bibinfo {author} {\bibfnamefont {T.}~\bibnamefont {Romanczukiewicz}},\ and\ \bibinfo {author} {\bibfnamefont {A.}~\bibnamefont {Wereszczynski}},\ }\href@noop {} {\bibfield  {journal} {\bibinfo  {journal} {Phys. Rev. D}\ }\textbf {\bibinfo {volume} {104}},\ \bibinfo {pages} {105022} (\bibinfo {year} {2021})}\BibitemShut {NoStop}%
\bibitem [{\citenamefont {Klinkhamer}\ and\ \citenamefont {Manton}(1984)}]{KM}%
  \BibitemOpen
  \bibfield  {author} {\bibinfo {author} {\bibfnamefont {F.~R.}\ \bibnamefont {Klinkhamer}}\ and\ \bibinfo {author} {\bibfnamefont {N.~S.}\ \bibnamefont {Manton}},\ }\href@noop {} {\bibfield  {journal} {\bibinfo  {journal} {Phys. Rev. D}\ }\textbf {\bibinfo {volume} {30}},\ \bibinfo {pages} {2212} (\bibinfo {year} {1984})}\BibitemShut {NoStop}%
\bibitem [{\citenamefont {Klinkhamer}(1993)}]{Klinkhamer}%
  \BibitemOpen
  \bibfield  {author} {\bibinfo {author} {\bibfnamefont {F.}~\bibnamefont {Klinkhamer}},\ }\href@noop {} {\bibfield  {journal} {\bibinfo  {journal} {Nucl. Phys. B}\ }\textbf {\bibinfo {volume} {410}},\ \bibinfo {pages} {343} (\bibinfo {year} {1993})}\BibitemShut {NoStop}%
\bibitem [{\citenamefont {James}(1992)}]{James}%
  \BibitemOpen
  \bibfield  {author} {\bibinfo {author} {\bibfnamefont {M.}~\bibnamefont {James}},\ }\href@noop {} {\bibfield  {journal} {\bibinfo  {journal} {Z. Phys. C}\ }\textbf {\bibinfo {volume} {55}},\ \bibinfo {pages} {515} (\bibinfo {year} {1992})}\BibitemShut {NoStop}%
\bibitem [{\citenamefont {Dashen}\ \emph {et~al.}(1975)\citenamefont {Dashen}, \citenamefont {Hasslacher},\ and\ \citenamefont {Neveu}}]{Dashen}%
  \BibitemOpen
  \bibfield  {author} {\bibinfo {author} {\bibfnamefont {R.~F.}\ \bibnamefont {Dashen}}, \bibinfo {author} {\bibfnamefont {B.}~\bibnamefont {Hasslacher}},\ and\ \bibinfo {author} {\bibfnamefont {A.}~\bibnamefont {Neveu}},\ }\href@noop {} {\bibfield  {journal} {\bibinfo  {journal} {Phys. Rev. D}\ }\textbf {\bibinfo {volume} {11}},\ \bibinfo {pages} {3424} (\bibinfo {year} {1975})}\BibitemShut {NoStop}%
\bibitem [{\citenamefont {Kudryavtsev}(1975)}]{Kudryavtsev}%
  \BibitemOpen
  \bibfield  {author} {\bibinfo {author} {\bibfnamefont {A.~E.}\ \bibnamefont {Kudryavtsev}},\ }\href@noop {} {\bibfield  {journal} {\bibinfo  {journal} {JETP Lett. (USSR) (Engl. Transl.)}\ }\textbf {\bibinfo {volume} {22}},\ \bibinfo {pages} {82} (\bibinfo {year} {1975})}\BibitemShut {NoStop}%
\bibitem [{\citenamefont {Bogolyubskii}\ and\ \citenamefont {Makhan’kov}(1976)}]{Bogolyubskii}%
  \BibitemOpen
  \bibfield  {author} {\bibinfo {author} {\bibfnamefont {I.~L.}\ \bibnamefont {Bogolyubskii}}\ and\ \bibinfo {author} {\bibfnamefont {V.~G.}\ \bibnamefont {Makhan’kov}},\ }\href@noop {} {\bibfield  {journal} {\bibinfo  {journal} {Pis’ma Zh. Eksp. Teor. Fiz.}\ }\textbf {\bibinfo {volume} {25}},\ \bibinfo {pages} {120} (\bibinfo {year} {1976})},\ \bibinfo {note} {[JETP Lett. \textbf{25}, 107 (1977)]}\BibitemShut {NoStop}%
\bibitem [{\citenamefont {Copeland}\ \emph {et~al.}(1995)\citenamefont {Copeland}, \citenamefont {Gleiser},\ and\ \citenamefont {Muller}}]{Copeland}%
  \BibitemOpen
  \bibfield  {author} {\bibinfo {author} {\bibfnamefont {E.}~\bibnamefont {Copeland}}, \bibinfo {author} {\bibfnamefont {M.}~\bibnamefont {Gleiser}},\ and\ \bibinfo {author} {\bibfnamefont {H.}~\bibnamefont {Muller}},\ }\href@noop {} {\bibfield  {journal} {\bibinfo  {journal} {Phys. Rev. D}\ }\textbf {\bibinfo {volume} {52}},\ \bibinfo {pages} {1920} (\bibinfo {year} {1995})}\BibitemShut {NoStop}%
\bibitem [{\citenamefont {Ollé}\ \emph {et~al.}(2020)\citenamefont {Ollé}, \citenamefont {Pujolàs},\ and\ \citenamefont {Rompineve}}]{Olle}%
  \BibitemOpen
  \bibfield  {author} {\bibinfo {author} {\bibfnamefont {J.}~\bibnamefont {Ollé}}, \bibinfo {author} {\bibfnamefont {O.}~\bibnamefont {Pujolàs}},\ and\ \bibinfo {author} {\bibfnamefont {F.}~\bibnamefont {Rompineve}},\ }\href@noop {} {\bibfield  {journal} {\bibinfo  {journal} {JCAP}\ }\textbf {\bibinfo {volume} {2020}}\bibinfo  {number} { (02)},\ \bibinfo {pages} {006}}\BibitemShut {NoStop}%
\bibitem [{\citenamefont {Kawasaki}\ \emph {et~al.}(2020)\citenamefont {Kawasaki}, \citenamefont {Nakano},\ and\ \citenamefont {Sonomoto}}]{Kawasaki}%
  \BibitemOpen
\bibfield  {number} {  }\bibfield  {author} {\bibinfo {author} {\bibfnamefont {M.}~\bibnamefont {Kawasaki}}, \bibinfo {author} {\bibfnamefont {W.}~\bibnamefont {Nakano}},\ and\ \bibinfo {author} {\bibfnamefont {E.}~\bibnamefont {Sonomoto}},\ }\href@noop {} {\bibfield  {journal} {\bibinfo  {journal} {JCAP}\ }\textbf {\bibinfo {volume} {2020}}\bibinfo  {number} { (01)},\ \bibinfo {pages} {047}}\BibitemShut {NoStop}%
\bibitem [{\citenamefont {Arvanitaki}\ \emph {et~al.}(2020)\citenamefont {Arvanitaki}, \citenamefont {Dimopoulos}, \citenamefont {Galanis}, \citenamefont {Lehner}, \citenamefont {Thompson},\ and\ \citenamefont {Tilburg}}]{Arvanitaki}%
  \BibitemOpen
\bibfield  {number} {  }\bibfield  {author} {\bibinfo {author} {\bibfnamefont {A.}~\bibnamefont {Arvanitaki}}, \bibinfo {author} {\bibfnamefont {S.}~\bibnamefont {Dimopoulos}}, \bibinfo {author} {\bibfnamefont {M.}~\bibnamefont {Galanis}}, \bibinfo {author} {\bibfnamefont {L.}~\bibnamefont {Lehner}}, \bibinfo {author} {\bibfnamefont {J.~O.}\ \bibnamefont {Thompson}},\ and\ \bibinfo {author} {\bibfnamefont {K.~V.}\ \bibnamefont {Tilburg}},\ }\href@noop {} {\bibfield  {journal} {\bibinfo  {journal} {Phys. Rev. D}\ }\textbf {\bibinfo {volume} {101}},\ \bibinfo {pages} {083014} (\bibinfo {year} {2020})}\BibitemShut {NoStop}%
\bibitem [{\citenamefont {Hindmarsh}\ and\ \citenamefont {Salmi}(2008)}]{Hindmarsh}%
  \BibitemOpen
  \bibfield  {author} {\bibinfo {author} {\bibfnamefont {M.}~\bibnamefont {Hindmarsh}}\ and\ \bibinfo {author} {\bibfnamefont {P.}~\bibnamefont {Salmi}},\ }\href@noop {} {\bibfield  {journal} {\bibinfo  {journal} {Phys. Rev. D}\ }\textbf {\bibinfo {volume} {77}},\ \bibinfo {pages} {105025} (\bibinfo {year} {2008})}\BibitemShut {NoStop}%
\bibitem [{\citenamefont {Gorghetto}\ \emph {et~al.}(2021)\citenamefont {Gorghetto}, \citenamefont {Hardy},\ and\ \citenamefont {Villadoro}}]{Gorghetto}%
  \BibitemOpen
  \bibfield  {author} {\bibinfo {author} {\bibfnamefont {M.}~\bibnamefont {Gorghetto}}, \bibinfo {author} {\bibfnamefont {E.}~\bibnamefont {Hardy}},\ and\ \bibinfo {author} {\bibfnamefont {G.}~\bibnamefont {Villadoro}},\ }\href@noop {} {\bibfield  {journal} {\bibinfo  {journal} {SciPost Phys.}\ }\textbf {\bibinfo {volume} {10}},\ \bibinfo {pages} {050} (\bibinfo {year} {2021})}\BibitemShut {NoStop}%
\bibitem [{\citenamefont {Blanco-Pillado}\ \emph {et~al.}(2021)\citenamefont {Blanco-Pillado}, \citenamefont {Jiménez-Aguilar},\ and\ \citenamefont {Urrestilla}}]{Blanco}%
  \BibitemOpen
  \bibfield  {author} {\bibinfo {author} {\bibfnamefont {J.~J.}\ \bibnamefont {Blanco-Pillado}}, \bibinfo {author} {\bibfnamefont {D.}~\bibnamefont {Jiménez-Aguilar}},\ and\ \bibinfo {author} {\bibfnamefont {J.}~\bibnamefont {Urrestilla}},\ }\href@noop {} {\bibfield  {journal} {\bibinfo  {journal} {JCAP}\ }\textbf {\bibinfo {volume} {2021}}\bibinfo  {number} { (01)},\ \bibinfo {pages} {027}}\BibitemShut {NoStop}%
\bibitem [{\citenamefont {Manton}\ and\ \citenamefont {Romanczukiewicz}(2023)}]{MR}%
  \BibitemOpen
\bibfield  {number} {  }\bibfield  {author} {\bibinfo {author} {\bibfnamefont {N.}~\bibnamefont {Manton}}\ and\ \bibinfo {author} {\bibfnamefont {T.}~\bibnamefont {Romanczukiewicz}},\ }\href@noop {} {\bibfield  {journal} {\bibinfo  {journal} {Phys. Rev. D}\ }\textbf {\bibinfo {volume} {107}},\ \bibinfo {pages} {085012} (\bibinfo {year} {2023})}\BibitemShut {NoStop}%
\bibitem [{\citenamefont {Oles}\ \emph {et~al.}(2023)\citenamefont {Oles}, \citenamefont {Queiruga}, \citenamefont {Romanczukiewicz},\ and\ \citenamefont {Wereszczynski}}]{Jose}%
  \BibitemOpen
  \bibfield  {author} {\bibinfo {author} {\bibfnamefont {K.}~\bibnamefont {Oles}}, \bibinfo {author} {\bibfnamefont {J.}~\bibnamefont {Queiruga}}, \bibinfo {author} {\bibfnamefont {T.}~\bibnamefont {Romanczukiewicz}},\ and\ \bibinfo {author} {\bibfnamefont {A.}~\bibnamefont {Wereszczynski}},\ }\href@noop {} {\bibfield  {journal} {\bibinfo  {journal} {Phys. Lett. B}\ }\textbf {\bibinfo {volume} {847}},\ \bibinfo {pages} {138300} (\bibinfo {year} {2023})}\BibitemShut {NoStop}%
\bibitem [{\citenamefont {Avelar}\ \emph {et~al.}(2008)\citenamefont {Avelar}, \citenamefont {Bazeia}, \citenamefont {Losano},\ and\ \citenamefont {Menezes}}]{Bazeia}%
  \BibitemOpen
  \bibfield  {author} {\bibinfo {author} {\bibfnamefont {A.~T.}\ \bibnamefont {Avelar}}, \bibinfo {author} {\bibfnamefont {D.}~\bibnamefont {Bazeia}}, \bibinfo {author} {\bibfnamefont {L.}~\bibnamefont {Losano}},\ and\ \bibinfo {author} {\bibfnamefont {R.}~\bibnamefont {Menezes}},\ }\href@noop {} {\bibfield  {journal} {\bibinfo  {journal} {Eur. Phys. J. C}\ }\textbf {\bibinfo {volume} {55}},\ \bibinfo {pages} {133} (\bibinfo {year} {2008})}\BibitemShut {NoStop}%
\bibitem [{\citenamefont {Alonso-Izquierdo}\ \emph {et~al.}(2023)\citenamefont {Alonso-Izquierdo}, \citenamefont {Navarro-Obregón}, \citenamefont {Oles}, \citenamefont {Queiruga}, \citenamefont {Romanczukiewicz},\ and\ \citenamefont {Wereszczynski}}]{Alberto}%
  \BibitemOpen
  \bibfield  {author} {\bibinfo {author} {\bibfnamefont {A.}~\bibnamefont {Alonso-Izquierdo}}, \bibinfo {author} {\bibfnamefont {S.}~\bibnamefont {Navarro-Obregón}}, \bibinfo {author} {\bibfnamefont {K.}~\bibnamefont {Oles}}, \bibinfo {author} {\bibfnamefont {J.}~\bibnamefont {Queiruga}}, \bibinfo {author} {\bibfnamefont {T.}~\bibnamefont {Romanczukiewicz}},\ and\ \bibinfo {author} {\bibfnamefont {A.}~\bibnamefont {Wereszczynski}},\ }\href@noop {} {\bibfield  {journal} {\bibinfo  {journal} {Phys. Rev. E}\ }\textbf {\bibinfo {volume} {108}},\ \bibinfo {pages} {064208} (\bibinfo {year} {2023})}\BibitemShut {NoStop}%
\bibitem [{\citenamefont {Manton}(2024)}]{Manton3}%
  \BibitemOpen
  \bibfield  {author} {\bibinfo {author} {\bibfnamefont {N.~S.}\ \bibnamefont {Manton}},\ }\href@noop {} {\bibfield  {journal} {\bibinfo  {journal} {J. Phys. A: Math. Theor.}\ }\textbf {\bibinfo {volume} {57}},\ \bibinfo {pages} {025202} (\bibinfo {year} {2024})}\BibitemShut {NoStop}%
\bibitem [{\citenamefont {Dorey}\ \emph {et~al.}(2011)\citenamefont {Dorey}, \citenamefont {Mersh}, \citenamefont {Romanczukiewicz},\ and\ \citenamefont {Shnir}}]{Dorey}%
  \BibitemOpen
  \bibfield  {author} {\bibinfo {author} {\bibfnamefont {P.}~\bibnamefont {Dorey}}, \bibinfo {author} {\bibfnamefont {K.}~\bibnamefont {Mersh}}, \bibinfo {author} {\bibfnamefont {T.}~\bibnamefont {Romanczukiewicz}},\ and\ \bibinfo {author} {\bibfnamefont {Y.}~\bibnamefont {Shnir}},\ }\href@noop {} {\bibfield  {journal} {\bibinfo  {journal} {Phys. Rev. Lett.}\ }\textbf {\bibinfo {volume} {107}},\ \bibinfo {pages} {091602} (\bibinfo {year} {2011})}\BibitemShut {NoStop}%
\bibitem [{\citenamefont {Adam}\ \emph {et~al.}(2022{\natexlab{a}})\citenamefont {Adam}, \citenamefont {Dorey}, \citenamefont {Martín-Caro}, \citenamefont {Huidobro}, \citenamefont {Oles}, \citenamefont {Romanczukiewicz}, \citenamefont {Shnir},\ and\ \citenamefont {Wereszczynski}}]{Garcia}%
  \BibitemOpen
  \bibfield  {author} {\bibinfo {author} {\bibfnamefont {C.}~\bibnamefont {Adam}}, \bibinfo {author} {\bibfnamefont {P.}~\bibnamefont {Dorey}}, \bibinfo {author} {\bibfnamefont {A.~G.}\ \bibnamefont {Martín-Caro}}, \bibinfo {author} {\bibfnamefont {M.}~\bibnamefont {Huidobro}}, \bibinfo {author} {\bibfnamefont {K.}~\bibnamefont {Oles}}, \bibinfo {author} {\bibfnamefont {T.}~\bibnamefont {Romanczukiewicz}}, \bibinfo {author} {\bibfnamefont {Y.}~\bibnamefont {Shnir}},\ and\ \bibinfo {author} {\bibfnamefont {A.}~\bibnamefont {Wereszczynski}},\ }\href@noop {} {\bibfield  {journal} {\bibinfo  {journal} {Phys. Rev. D}\ }\textbf {\bibinfo {volume} {106}},\ \bibinfo {pages} {125003} (\bibinfo {year} {2022}{\natexlab{a}})}\BibitemShut {NoStop}%
\bibitem [{\citenamefont {Ghersi}\ and\ \citenamefont {Braden}(2023)}]{Ghersi}%
  \BibitemOpen
  \bibfield  {author} {\bibinfo {author} {\bibfnamefont {J.~T.~G.}\ \bibnamefont {Ghersi}}\ and\ \bibinfo {author} {\bibfnamefont {J.~N.}\ \bibnamefont {Braden}},\ }\href@noop {} {\bibfield  {journal} {\bibinfo  {journal} {Phys. Rev. D}\ }\textbf {\bibinfo {volume} {108}},\ \bibinfo {pages} {096017} (\bibinfo {year} {2023})}\BibitemShut {NoStop}%
\bibitem [{\citenamefont {Blaschke}\ \emph {et~al.}(2024)\citenamefont {Blaschke}, \citenamefont {Roma\'nczukiewicz}, \citenamefont {S\l{}awi\'nska},\ and\ \citenamefont {Wereszczy\'nski}}]{Blaschke:2024uec}%
  \BibitemOpen
  \bibfield  {author} {\bibinfo {author} {\bibfnamefont {F.}~\bibnamefont {Blaschke}}, \bibinfo {author} {\bibfnamefont {T.}~\bibnamefont {Roma\'nczukiewicz}}, \bibinfo {author} {\bibfnamefont {K.}~\bibnamefont {S\l{}awi\'nska}},\ and\ \bibinfo {author} {\bibfnamefont {A.}~\bibnamefont {Wereszczy\'nski}},\ }\href@noop {} {\  (\bibinfo {year} {2024})},\ \Eprint {https://arxiv.org/abs/2403.00443} {arXiv:2403.00443 [hep-th]} \BibitemShut {NoStop}%
\bibitem [{\citenamefont {Manton}\ \emph {et~al.}(2021)\citenamefont {Manton}, \citenamefont {Oleś}, \citenamefont {Romańczukiewicz},\ and\ \citenamefont {Wereszczyński}}]{Oles}%
  \BibitemOpen
  \bibfield  {author} {\bibinfo {author} {\bibfnamefont {N.}~\bibnamefont {Manton}}, \bibinfo {author} {\bibfnamefont {K.}~\bibnamefont {Oleś}}, \bibinfo {author} {\bibfnamefont {T.}~\bibnamefont {Romańczukiewicz}},\ and\ \bibinfo {author} {\bibfnamefont {A.}~\bibnamefont {Wereszczyński}},\ }\href@noop {} {\bibfield  {journal} {\bibinfo  {journal} {Phys. Rev. D}\ }\textbf {\bibinfo {volume} {103}},\ \bibinfo {pages} {025024} (\bibinfo {year} {2021})}\BibitemShut {NoStop}%
\bibitem [{\citenamefont {Adam}\ \emph {et~al.}(2022{\natexlab{b}})\citenamefont {Adam}, \citenamefont {Manton}, \citenamefont {Oles}, \citenamefont {Romanczukiewicz},\ and\ \citenamefont {Wereszczynski}}]{Andrzej}%
  \BibitemOpen
  \bibfield  {author} {\bibinfo {author} {\bibfnamefont {C.}~\bibnamefont {Adam}}, \bibinfo {author} {\bibfnamefont {N.}~\bibnamefont {Manton}}, \bibinfo {author} {\bibfnamefont {K.}~\bibnamefont {Oles}}, \bibinfo {author} {\bibfnamefont {T.}~\bibnamefont {Romanczukiewicz}},\ and\ \bibinfo {author} {\bibfnamefont {A.}~\bibnamefont {Wereszczynski}},\ }\href@noop {} {\bibfield  {journal} {\bibinfo  {journal} {Phys. Rev. D}\ }\textbf {\bibinfo {volume} {105}},\ \bibinfo {pages} {065012} (\bibinfo {year} {2022}{\natexlab{b}})}\BibitemShut {NoStop}%
\bibitem [{\citenamefont {Navarro-Obregón}\ \emph {et~al.}(2023)\citenamefont {Navarro-Obregón}, \citenamefont {Nieto},\ and\ \citenamefont {Queiruga}}]{Navarro}%
  \BibitemOpen
  \bibfield  {author} {\bibinfo {author} {\bibfnamefont {S.}~\bibnamefont {Navarro-Obregón}}, \bibinfo {author} {\bibfnamefont {L.}~\bibnamefont {Nieto}},\ and\ \bibinfo {author} {\bibfnamefont {J.}~\bibnamefont {Queiruga}},\ }\href@noop {} {\bibfield  {journal} {\bibinfo  {journal} {Phys. Rev. E}\ }\textbf {\bibinfo {volume} {108}},\ \bibinfo {pages} {044216} (\bibinfo {year} {2023})}\BibitemShut {NoStop}%
\bibitem [{\citenamefont {Fodor}\ \emph {et~al.}(2008{\natexlab{a}})\citenamefont {Fodor}, \citenamefont {Forgacs}, \citenamefont {Horvath},\ and\ \citenamefont {Lukacs}}]{Fodor:2008es}%
  \BibitemOpen
  \bibfield  {author} {\bibinfo {author} {\bibfnamefont {G.}~\bibnamefont {Fodor}}, \bibinfo {author} {\bibfnamefont {P.}~\bibnamefont {Forgacs}}, \bibinfo {author} {\bibfnamefont {Z.}~\bibnamefont {Horvath}},\ and\ \bibinfo {author} {\bibfnamefont {A.}~\bibnamefont {Lukacs}},\ }\href {https://doi.org/10.1103/PhysRevD.78.025003} {\bibfield  {journal} {\bibinfo  {journal} {Phys. Rev. D}\ }\textbf {\bibinfo {volume} {78}},\ \bibinfo {pages} {025003} (\bibinfo {year} {2008}{\natexlab{a}})}\BibitemShut {NoStop}%
\bibitem [{\citenamefont {Fodor}\ \emph {et~al.}(2008{\natexlab{b}})\citenamefont {Fodor}, \citenamefont {Forgács}, \citenamefont {Horváth},\ and\ \citenamefont {Lukács}}]{Fodor}%
  \BibitemOpen
  \bibfield  {author} {\bibinfo {author} {\bibfnamefont {G.}~\bibnamefont {Fodor}}, \bibinfo {author} {\bibfnamefont {P.}~\bibnamefont {Forgács}}, \bibinfo {author} {\bibfnamefont {Z.}~\bibnamefont {Horváth}},\ and\ \bibinfo {author} {\bibfnamefont {A.}~\bibnamefont {Lukács}},\ }\href@noop {} {\bibfield  {journal} {\bibinfo  {journal} {Phys. Rev. D}\ }\textbf {\bibinfo {volume} {78}},\ \bibinfo {pages} {025003} (\bibinfo {year} {2008}{\natexlab{b}})}\BibitemShut {NoStop}%
\bibitem [{\citenamefont {Adam}\ \emph {et~al.}(2020)\citenamefont {Adam}, \citenamefont {Oles}, \citenamefont {Romanczukiewicz},\ and\ \citenamefont {Wereszczynski}}]{Adam:2019uat}%
  \BibitemOpen
  \bibfield  {author} {\bibinfo {author} {\bibfnamefont {C.}~\bibnamefont {Adam}}, \bibinfo {author} {\bibfnamefont {K.}~\bibnamefont {Oles}}, \bibinfo {author} {\bibfnamefont {T.}~\bibnamefont {Romanczukiewicz}},\ and\ \bibinfo {author} {\bibfnamefont {A.}~\bibnamefont {Wereszczynski}},\ }\href {https://doi.org/10.1103/PhysRevE.102.062214} {\bibfield  {journal} {\bibinfo  {journal} {Phys. Rev. E}\ }\textbf {\bibinfo {volume} {102}},\ \bibinfo {pages} {062214} (\bibinfo {year} {2020})}\BibitemShut {NoStop}%
\bibitem [{\citenamefont {Barashenkov}\ and\ \citenamefont {Oxtoby}(2009)}]{Oxtoby}%
  \BibitemOpen
  \bibfield  {author} {\bibinfo {author} {\bibfnamefont {I.~V.}\ \bibnamefont {Barashenkov}}\ and\ \bibinfo {author} {\bibfnamefont {O.~F.}\ \bibnamefont {Oxtoby}},\ }\href@noop {} {\bibfield  {journal} {\bibinfo  {journal} {Phys. Rev. E}\ }\textbf {\bibinfo {volume} {80}},\ \bibinfo {pages} {026608} (\bibinfo {year} {2009})}\BibitemShut {NoStop}%
\end{thebibliography}%

\end{document}